# Aggregating Epigenetic Clocks to Study Human Capital Formation[*]

## Giorgia Menta[1,2], Pietro Biroli[3], Divya Mehta[4],

## Conchita D'Ambrosio[2], Deborah Cobb-Clark[5]

*[1]Luxembourg Institute of Socio-Economic Research (LISER), [2]University of Luxembourg,*

*[3]University of Bologna, [4]Queensland University of Technology, [5]University of Sydney*

This version: August 28, 2025

## Abstract

Epigenetics is the study of how people's behavior and environments influence the way their genes are expressed, even though their DNA sequence is itself unchanged. By aggregating age-related epigenetic markers, epigenetic 'clocks' have become the leading tool for studying biological aging. We make an important contribution by developing a novel, integrated measure of epigenetic aging – the Multi EpiGenetic Age (MEGA) clock – which combines several existing epigenetic clocks to reduce measurement error and improve estimation efficiency. We use the MEGA clock in three empirical contexts to show that: i) accelerated epigenetic aging in adolescence is associated with worse educational, mental-health, and labor market outcomes in early adulthood; ii) exposure to child maltreatment before adolescence is associated with half a year higher epigenetic aging; and iii) that entering school one year later accelerates epigenetic aging by age seven, particularly among disadvantaged children. The MEGA clock is robust to alternative methods for constructing it, providing a flexible and interpretable approach for incorporating epigenetic data into a wide variety of settings.

[*]We would like to thank Nancy Kong and Anita Sathyanarayanan for their input in establishing the conceptual underpinnings of the project. Tiger Mathieson and Sura Majeed provided excellent research assistance. We thank seminar and conference participants at the University of Bristol, the University of Essen, NTU Singapore, Chulalongkorn University, the 16[th] Annual IGSS Conference, the 38[th] ESPE Annual Conference, and the 2024 AASLE Conference for useful comments and suggestions. We are grateful to all the families who took part in this study, the midwives for their help in recruiting them, and the whole ALSPAC team, which includes interviewers, computer and laboratory technicians, clerical workers, research scientists, volunteers, managers, receptionists and nurses. The U.K. Medical Research Council and Welcome (Grant ref: 217065/Z/19/Z) and the University of Bristol provide core support for ALSPAC. This publication is the work of the authors and all authors will serve as guarantors for the contents of this paper. A comprehensive list of grants funding the ALSPAC project is available on their website (http://www.bristol.ac.uk/alspac/external/documents/grant-acknowledgements.pdf). The authors are grateful for research support from the Australian Research Council (ARC) (grant numbers DP200100979, CE200100025) and from the Fonds National de la Recherche Luxembourg (Grant C19/SC/13650569/ALAC).

## 1. Introduction

Nature and nurture form a two-way street: human biology and environmental conditions influence people's life chances through a dynamic interplay. Our genetic code affects the environments we grow up in and the trajectories we follow (Plomin and Bergeman, 1991; Cole 2009; Belsky and Harden, 2019). External shocks and policy interventions, in turn, can alter biological processes in ways that persist across the life course (Hertzman, 1999; Shonkoff *et al.*, 2009). In economics, early evidence from twin and adoption studies (e.g. Leibowitz, 1974; Goldberger, 1976; Taubman, 1976; Beauchamp *et al.*, 2017; Björklund *et al.*, 2006; Cesarini *et al.*, 2010) has contributed to recent advances in behavioral genetics (e.g. Okbay *et al.*, 2022; Abdellaoui *et al.*, 2023; Benjamin *et al.*, 2007, 2024) and the integration of genes into economic models (e.g. Papageorge and Thom, 2020; Houmark *et al.*, 2024; Sanz-de-Galdeano and Terskaya, 2025). While genetic variants are fixed at conception, genetic expression is not. Epigenetics – the study of changes in gene expression that occur without alterations to the DNA sequence – has opened a new frontier for understanding the process through which people's environments affect their biological functioning. Epigenetic modifications can happen through exposure to adverse environments (Cole, 2009); leave a lasting biological footprint (Szyf, 2009); and shape economic choices (Almond and Currie, 2011). Social scientists are increasing their efforts to understand the way socio-economic conditions shape the epigenome; yet, with a few recent exceptions (Schmitz and Duque, 2022; Costi *et al.* 2025a, 2025b), the economics literature on epigenetics remains relatively limited. This is unfortunate because measuring, understanding, and causally estimating these epigenetic influences can provide insights into the pathways through which economic policies and social environments affect public health and productivity.

We contribute to this emerging field by introducing a novel measure of epigenetic aging and showing how it can be systematically incorporated into economic models to better understand the dynamics of health and human capital formation. Epigenetic aging refers to the pace at which the body ages biologically and is based on changes in gene expression rather than the passage of time. This process is typically captured using epigenetic 'clocks', which employ machine learning algorithms to aggregate DNA methylation data into empirically quantifiable measures of biological age.[1] Biological age has been shown to be a more reliable predictor of physiological functioning, disease risk, health outcomes, and mortality than chronological age (Kotschy *et al.*, 2025). By summarizing complex epigenetic information, epigenetic clocks offer a useful and quantifiable measure of biological aging that is valuable in both biological and social-science research. The growing number of available clocks, however, can make interpretation challenging (Bell *et al.*, 2019; Teschendorff and Horvath, 2025). Intended to capture the same latent concept – epigenetic aging – the algorithms used to define different epigenetic clocks nevertheless diverge in terms of methodology and selection of relevant epigenetic sites, leading to conceptual differences across clocks as well as measurement error. Applied researchers are often left without clear guidance on which epigenetic measure to use. Moreover, while technological advances have made it less costly to include epigenetic markers in social-science data, sample sizes often remain small, limiting statistical power.

We make four key contributions, one methodological and three empirical. The first is to propose a methodological approach for combining existing epigenetic clocks into a single metric – the Multi EpiGenetic Age (MEGA) clock, which reduces measurement error and provides a more efficient and robust measure of latent epigenetic age. Our strategy integrates several well-established clocks (our empirical application uses four: Horvath, 2013; Hannum *et al.*, 2013; Levine *et al.*, 2018; Lu *et al.*, 2019), each trained to predict different aspects of aging and health, into a unified measure. We enhance the robustness of this approach by leveraging three alternative empirical methods for combining clocks: (i) exploratory factor analysis; (ii) weighted indexing; and (iii) structural equation modeling. Using epigenetic and behavioral data from a large birth cohort in the U.K., the Avon Longitudinal Study of Parents and Children (ALSPAC), we demonstrate that the MEGA clock reduces the measurement error inherent in individual epigenetic clocks, providing more precise estimates of the determinants and consequences of epigenetic aging. The MEGA clock also retains the intuitive age-scale interpretation of existing epigenetic clocks, making it directly comparable to chronological age. Given the growing availability of pre-computed epigenetic clocks in many large-scale data sets (e.g. U.K. Understanding Society, U.S. Health and Retirement Study, U.S. National Longitudinal Study of Adolescent to Adult Health,

---

[1] DNA methylation is a chemical process in which a small molecule, called a methyl group, is added to a DNA molecule. This addition can affect how genes are expressed, effectively turning them "on" or "off" which has implications for disease risk and long-term biological outcomes. DNA methylation is the most studied form of epigenetic modifications.

U.S. The Future of Families and Child Wellbeing Study), the MEGA clock is easy to implement without requiring in-depth biological expertise. Importantly, our methodological approach is designed to accommodate future developments – as new epigenetic clocks are developed, they can be readily incorporated, provided they capture the same latent construct of epigenetic aging.

Our three empirical contributions leverage the ALSPAC data to illustrate how the MEGA clock can be incorporated into both observational and causal analyses of human capital formation across the life cycle. In our first application, we assess the usefulness of the MEGA clock by investigating the association between age acceleration and human capital later in life. Faster epigenetic aging has been shown to be predictive of numerous outcomes, including mortality (e.g. Faul *et al.*, 2023; Lu *et al.*, 2019; Perna *et al.*, 2016) and disease incidence (e.g. Horvath, 2013; McCartney *et al.*, 2018). The biological processes linked to accelerated epigenetic aging – which include, among others, the regulation of the immune system, lipid function, and neuronal pathways (Han *et al.*, 2018; Lu *et al.*, 2019) – may well be detrimental to the formation of cognitive and socio-emotional skills, as suggested by studies looking at epigenetic aging and cognitive performance in children and young adults (e.g. Niccodemi *et al.*, 2022; Costi *et al.*, 2025a). We conceptualize the MEGA clock as a potential input in the human capital production function, hypothesizing that adolescents who experience accelerated epigenetic aging will exhibit worse cognitive and socio-emotional outcomes in early adulthood, even after controlling for family background, cognitive and socio-emotional skills, and health. Our results confirm that the MEGA clock is an efficient measure of biological aging and suggest that epigenetic age acceleration in adolescence predicts fewer years of education, worse mental health, and weaker attachment to the labor market, above and beyond traditional health biomarkers (e.g. BMI) and health behaviors (e.g. smoking and drinking).

We next turn to consideration of the shocks that might influence epigenetic aging. We begin by examining the association of exposure to child abuse on epigenetic age acceleration in late adolescence.[2] Child maltreatment is of particular interest because it is an external, child-specific experience with well-documented adverse consequences (Currie and Spatz Widom, 2010; Fletcher, 2009; Henkaus, 2022; Suglia *et al.*, 2014) and has been linked to widespread modifications across the epigenome (Lawn *et al.*, 2018). We contribute to this literature by shifting the focus from general epigenetic changes to epigenetic aging, showing how experiences of child abuse are associated with accelerated epigenetic aging when we adopt a longitudinal approach that accounts for different abuse trajectories and critical developmental periods (Heckman, 2007; Cunha *et al.*, 2010). Our results show that exposure to child maltreatment before adolescence is associated with half a year of accelerated epigenetic aging, a result comparable in magnitude to the age acceleration observed for children born into socio-economic disadvantage (see Fiorito *et al.*, 2017).

---

[2] Age acceleration occurs when an individual's epigenetic age (here, the MEGA clock) outpaces their chronological age, signaling a faster rate of biological aging relative to the passage of time.

In our final empirical study, we provide causal evidence of the biological consequences of early school exposure by examining how the timing of school entry influences epigenetic aging. Using a sharp regression discontinuity design (RDD) around the U.K. school-entry age cutoff, we identify the causal impact of entering school one year later on epigenetic age acceleration at age seven. This approach, widely used in economics to estimate the effects of school starting age on cognitive and labor market outcomes (e.g. Bedard and Dhuey, 2006; Black, Devereux, and Salvanes, 2011; Fredriksson and Öckert, 2014), allows us to be the first to isolate the biological consequences of delayed school entry. We find that children born just after the cutoff – who start school one year later – exhibit faster epigenetic aging (0.5 – 1.2 years) than those who start earlier, with especially pronounced effects among children from lower socio-economic backgrounds. These findings suggest that early educational environments may play a protective role in shaping biological development, potentially by promoting healthier routines and buffering stress exposure (Anderson *et al.*, 2011; Holford and Rabe, 2022). By embedding an RDD within an epigenetic framework, we show that institutional factors can causally affect epigenetic aging trajectories as early as mid-childhood.

Our research speaks to several literatures. First, we advance the emerging inter-disciplinary literature that uses epigenetic clocks by proposing a methodological innovation that addresses key challenges faced by researchers. Our approach provides a simple and scalable method to reduce reliance on arbitrary clock selection, improve the replicability of results across different clocks, and increase statistical power without expanding sample size. To our knowledge, this is the first study to build an integrated epigenetic clock based on other existing clocks. Recently, Martinez *et al.* (2025) have proposed a complementary, yet distinct, approach. They use a battery of biomarkers, including epigenetic clocks, and confirmatory factor analysis, to identify three factors they interpret as epigenetic age, systemic biological age, and immune age.

Second, we are the first to estimate the causal effect of age at school-entry on health and biological outcomes, and among the first to estimate the causal effect of quasi-exogenous shocks or policies on epigenetic aging (see Schmitz and Duque, 2022 for an exception).[3] We also contribute to the rich observational literature linking abuse with epigenetic aging, incorporating measures taken across different developmental periods and perspectives (i.e., from mothers, fathers, and children), and adding a temporal dimension to the analysis (see Cecil *et al.*, 2020, and Rubens *et al.*, 2023, for systematic reviews). Using data from ALSPAC mothers, Lawn *et al.* (2018) find that sexual abuse experienced by the age of 17 is associated with an epigenetic age that is 3.4 year higher in adulthood (Horvath clock), after adjusting for childhood and adulthood socio-economic positioning. Marini *et al.* (2020) look at early childhood DNA methylation in ALSPAC children and find that sexual and physical abuse, especially during early and middle-childhood sensitive periods, are associated with accelerated epigenetic aging at age seven. The effect is driven by girls and is only found when using the Hannum rather than the Horvath's clock. Lussier *et al.* (2023) also adopt a longitudinal approach, examining broad changes to the epigenome associated with abuse. Our focus on the portion of the epigenome that predicts aging, disease and mortality (as captured by the MEGA clock) contributes to a more nuanced understanding of the epigenetic footprint of child adversity.

Third, our research supports the growing evidence that epigenetic age is a valid and policy-relevant measure of health that can serve as a feasible short-term endpoint for evaluating the impact of interventions aimed at improving well-being. Age acceleration has been shown to be predictive of all-cause mortality (Faul *et al.*, 2023; Lu *et al.*, 2019; Perna *et al.*, 2016), healthcare utilization (Davillas and Jones, 2024), and the occurrence of diseases such as cancer (Horvath, 2013; Dugué *et al.*, 2018; Perna *et al.*, 2016) and Alzheimer's disease (McCartney *et al.*, 2018). In children and young adults, age acceleration has been shown to be negatively associated with test-scores (Niccodemi *et al.*, 2022), cognitive functioning (Raffington *et al.*, 2023a; Costi *et al.*, 2025a), and to increase social disparities in mental health (Raffington *et al.*, 2023b). Our focus on young adults contributes to the understanding of the early-adulthood consequences of accelerated epigenetic aging in the domains of health and labor market outcomes.

---

[3] A small clinical literature, typically based on samples of a few dozen participants, has examined the short-term biological-aging effects of randomized interventions aimed at improving diet and physical activity (e.g. Chen *et al.*, 2019; Fitzgerald *et al.*, 2021; Nwanaji *et al.*, 2021).

Finally, our life-course approach maps out experiences both in childhood and adolescence, providing a comprehensive view of the way that early-life conditions influence epigenetic aging and subsequent outcomes. This holistic perspective bridges gaps in this literature (e.g. Korous *et al.*, 2023; Petrovic *et al.*, 2023), offering new insights into the interplay between environmental factors and biological processes over time.

## 2. The Conceptual Underpinnings of Epigenetic Clocks

All humans age, but the rate at which we do so differs considerably from person to person. Chronological age measures how long someone has lived, while biological age captures the state of people's physiological and cellular systems based on various physiological and molecular markers. Biological aging is associated with progressive loss of function at the cellular, tissue, and organ levels, further accelerating the general decline in physical functioning and cognitive performance (López-Otín *et al.*, 2013). We are interested in a person's biological age because it is a more accurate measure of their functional capacity than is their chronological age.

But how can one measure biological aging? One well-studied process underlying biological aging is epigenetics, that is, the ensemble of reversible chemical and structural alterations to the genome that can lead to long-term changes in gene activity even though the underlying DNA sequence itself has not been altered (Klengel *et al.*, 2014). The most studied form of epigenetics is DNA methylation.[4] DNA methylation happens when a methyl group binds to a CpG site – a particular section of the DNA sequence where a cytosine nucleotide (C) is adjacent to a guanine nucleotide (G), with one phosphate (p) in between – playing a critical role in regulating gene expression by essentially turning genes "on" or "off". DNA methylation patterns are relatively stable over time and can be measured reliably at scale in large cohort studies, allowing researchers to link biological regulation to health and human capital outcomes across the life course.

---

[4] While other forms of epigenetic changes exist (e.g. histone modification), current biotechnologies have become increasingly cost-efficient in the stabilization and extraction of DNA methylation from human tissues. As a result, there is a wealth of data and well-characterized tools available for studying DNA methylation.

Research has shown that, as people age, certain parts of their DNA become systematically more or less methylated. This insight laid the foundation for the development of epigenetic biomarkers of aging – referred to as epigenetic clocks – that are designed to predict either chronological aging or physiological decline using DNA methylation levels. Epigenetic clocks are constructed using supervised machine-learning algorithms (e.g. penalized elastic net regression) to address the high-dimensionality of the data and prevent overfitting, since epigenetic datasets typically contain substantially fewer individuals than measured CpG sites. The algorithms usually consist of two phases: development and prediction. In the development phase, the parameters of the algorithm are adjusted to achieve the greatest prediction accuracy in large training samples. These estimated parameters are then used in prediction (estimation) samples to construct epigenetic clocks as weighted scores of DNA methylation at selected CpG sites, using the weights learned from the training sample used in model development. Adopting the same general approach, existing epigenetic clocks differ along four main dimensions: (i) which CpG sites are included in the training sample and the resulting weights identified by the algorithms; (ii) the type of tissue used to measure DNA methylation (e.g. blood, saliva); (iii) the outcome they are meant to predict (e.g., chronological age, physical fitness, mortality) and (iv) the sample that they are trained on.

## 2.1. First-Generation Clocks

The first two widely used epigenetic clocks, Horvath's DNA methylation age clock and Hannum's clock, were developed in 2013 to predict chronological age using DNA methylation data. The Horvath clock captures age-related DNA methylation patterns, using the combined methylation status of 353 CpG sites. It is a multi-organ clock that predicts chronological age from embryo to old age using DNA methylation data on many tissues and organs (e.g. whole blood, cerebellum, colon, kidney, liver, lungs). The Hannum clock similarly uses elastic net regression to predict people's chronological age, identifying 71 CpG sites that can be used to accurately predict age (Hannum *et al.*, 2013). Epigenetic age as estimated by Hannum's clock has been shown to be a more accurate predictor of chronological age than Horvath's clock when adult blood samples are used, since this is the tissue on which the model was trained. Estimates of Hannum epigenetic age have been shown to be biased, however, when applied in non-blood tissues (Simpkin *et al.*, 2016) and in children (Hannum *et al.*, 2013).

Because the algorithms underlying these clocks were trained on chronological age, critics have argued that these "first-generation" clocks may, in fact, exclude CpGs whose methylation patterns reflect biological age variation. As a result of being designed to detect age-dependent patterns, these algorithms can achieve remarkable chronological age prediction. Clocks that achieve almost perfect chronological age prediction, however, have been shown to be worse predictors of mortality (Zhang *et al.*, 2019).

## 2.2. Second-Generation Clocks

In recent years, a second generation of DNA methylation-based biomarkers for aging has emerged in which CpGs associated with organ-system functioning are captured alongside chronological age. In 2018, Levine developed the PhenoAge clock, trained to predict not only chronological age but other aging-related indicators (e.g. blood glucose, liver and kidney markers of function).[5] This clock, while also a good predictor of chronological age, has been extensively validated; it can predict age-related health outcomes more effectively than the first-generation clocks and can differentiate morbidity and mortality risks among individuals of identical chronological age (Levine *et al.*, 2018). Subsequently, Horvath developed GrimAge which surpassed previous clocks in predicting both age-related disease as well as mortality. GrimAge is a linear combination of DNA methylation-based surrogate biomarkers for health-related plasma proteins, smoking pack-years, sex, and chronological age. It has been shown to be a stronger predictor of lifespan, age-related conditions, disease, and mortality risk compared to the widely used Horvath's clock (Lu *et al.*, 2019).

The number of CpGs selected by the algorithms underlying epigenetic clocks vary substantially across different models. The Horvath clock incorporates 353 age-related CpGs, while the Hannum clock uses 71 CpGs, PhenoAge employs 513 CpGs, and GrimAge includes 1,030 CpGs. Beyond these numerical differences, the specific CpG sites selected for each model also show limited overlap. The PhenoAge clock shares only 41 CpGs with the Horvath clock, while both PhenoAge and Horvath clocks have merely 5 CpGs in common with the Hannum clock (Levine *et al.*, 2018).[6]

---

[5] The model combined 10 clinical characteristics, including chronological age, albumin, creatinine, glucose, C-reactive protein levels, lymphocyte percentage, mean cell volume, red blood cell distribution width, alkaline phosphatase, and white blood cell count. Based on 513 age-related CPGs on 3 chips (27 K, 450 K, 850 K), the DNA methylation PhenoAge has achieved greater applicability across chip platforms than other clocks.

## 2.3.  From Epigenetic Age to Age Acceleration

Epigenetic age, as estimated by DNA methylation-based clocks, provides a biological measure of aging that often differs from chronological age. The difference between a person's predicted epigenetic age and their actual chronological age is commonly referred to as age acceleration, typically measured using the residuals from a regression of epigenetic age on chronological age. Understanding age acceleration is important because it provides insights into individual-specific variation in the aging process that cannot be gained by simply counting years lived. People of the same chronological age often differ substantially in their biological condition due to genetics, lifestyle, and environmental factors. Age acceleration highlights who is aging faster or slower than average, providing a personalized metric (biomarker) that better predicts health risks, disease progression, and life outcomes.

In all empirical applications, we control for chronological age, allowing estimates of the determinants and consequences of epigenetic age to be interpreted directly as age acceleration (see Section 5).

## 3.  From Several Clocks to One Latent Construct: the MEGA Clock

Different epigenetic clocks capture different aspects of the biological processes linked to aging, disease onset and all-cause mortality. The substantial differences in training methodologies, CpG site selection, and construction approaches create challenges for researchers seeking to determine the most appropriate epigenetic aging measure for their specific analysis. Drawing from the social-sciences literature on the measurement of latent factors, we propose a novel way of harnessing the information coming from different epigenetic clocks into a unique measure of epigenetic aging which we name Multi EpiGenetic Age (MEGA) clock. Although our approach can be implemented with several different clocks that capture the same underlying construct, our current focus is on four clocks: two first-generation clocks (the Horvath and the Hannum clocks), which have been trained to predict chronological age; and two second-generation clocks (the PhenoAge and GrimAge clocks), which have been trained to predict lifespan (functional stage). Our rationale for this choice is twofold. On the one hand, the robustness and replicability of these clocks have been widely shown across a variety of samples and tissues (Lu *et al.*, 2019; Maddock *et al.*, 2020; McCrory *et al.*, 2021). On the other hand, these clocks all rely on genome-wide DNA methylation data derived from the same methylation profiling technology (i.e. Illumina Infinium arrays)[7] and are constructed using similar methods (i.e. penalized elastic net algorithms). If different epigenetic clocks truly capture different facets of the molecular physiological determinants of aging and ill health, it is reasonable to think of these clocks as separate indicators of the same latent concept. Combining the information contained in these four clocks would then be expected to reduce measurement error, leading to a more robust, holistic measure of DNA methylation age.

---

[6] Models that include a large number of CpG sites tend to be more robust and accurate than those with fewer (Liu *et al.*, 2020; Lu *et al.*, 2019). However, epigenetic clocks aim to capture broad features of the methylome, and Horvath and Raj (2018) show that including a moderate number of age-associated CpGs is sufficient to yield highly reliable models.

We construct three versions of the MEGA clock, each based on a different methodological approach with varying underlying identification assumptions. In the empirical applications we will rely on the four clocks defined above (Horvath, Hannum, PhenoAge, GrimAge), however, it is important to note that the procedures described below are generalizable to any set of epigenetic clocks. The first MEGA clock, $MEGA_{WGT}$, relies on a weighted index approach proposed by Anderson (2008). Here, clocks are combined using a weighted sum with the weights equal to the row-sums of the inverse variance-covariance matrix of the clocks. Specifically, let $\{C_1, \ldots, C_K\}$ be a set of epigenetic clocks, all expressed in the same unit (years of age), and $M$ be their variance-covariance matrix. Further, $I_K$ is a vector of ones with length $K$. The $MEGA_{WGT}$ clock is defined as:

$$MEGA_{WGT} = \frac{\sum_{k=1}^{K} w_{k,1} C_k}{\sum_{k=1}^{K} w_{k,1}} \tag{1}$$

where $w_{k,1}$ is the $k$-th element of vector $w = M^{-1} I_K$. As described in Anderson (2008), using the $MEGA_{WGT}$ clock constructed with this weighting procedure in a regression is analogous to the joint estimation of seemingly unrelated regression, one for each individual clock, constraining all clock-related coefficients to be equal, and corresponds to an efficient Generalized Least Squares (GLS) estimator.

---

[7] Illumina Infinium arrays are a profiling technology that measures DNA methylation. For example, Illumina Infinium 450k arrays measure DNA methylation in up to 450k CpG sites.

The second MEGA clock, $MEGA_{FA}$, relies instead on exploratory factor analysis, which is commonly used to identify one or more latent constructs using a series of measures. It is a data-driven method that aims to explain the observed variability in the data by grouping variables that tend to co-vary and extracting a smaller number of latent factors that account for the patterns in the observed variables. From a theoretical standpoint, since $\{C_1, \dots, C_K\}$ are selected to capture the same latent concept of epigenetic age, we expect only one factor to be retained by the exploratory factor analysis. We interpret this factor as the $MEGA_{FA}$. Let $L_K$ be the vector of factor loadings for each of the $K$ clocks. Then $MEGA_{FA}$ is defined similarly to $MEGA_{WGT}$ as:

$$MEGA_{FA} = \frac{\sum_{k=1}^{K} u_{k,1} C_k}{\sum_{k=1}^{K} u_{k,1}} \qquad (2)$$

where $u_{k,1}$ is the $k$-th element of vector $u = M^{-1} L_K$. Both $MEGA_{WGT}$ and $MEGA_{FA}$ rely on the clocks' variance-covariance matrix $M$, but they do so in different ways. $MEGA_{WGT}$ downweights redundant clocks and gives more weight to those adding independent information, whereas $MEGA_{FA}$ additionally emphasizes the common variance across clocks, as captured by their factor loadings, thereby extracting the shared aging signal.

Lastly, $MEGA_{SEM}$, uses Structural Equation Modelling and relies on the strongest identification assumptions. The advantage of Structural Equation Modelling is that it allows us to simultaneously model both observed and latent variables and their direct relationships, as well as error terms. Informed by the factor analysis, the $MEGA_{SEM}$ clock is estimated as the latent factor of a system of measurement equations based on the individual clocks and a behavioral (or structural) equation, simultaneously modelling its relationship with other shocks or outcomes – such as child abuse, age at school-entry, or labor market outcomes. The system of measurement equations can be illustrated as follows:

$$C_k = \lambda_k \, EA^* + \varepsilon_k, \qquad k = 1, \dots, K \qquad (3)$$

where $C_k$, the $k$-th epigenetic clock, can be seen as an indicator of latent epigenetic age, denoted $EA^*$. $\lambda_k$ is the factor loading parameter and $\varepsilon_k$ is a stochastic error term.

In addition, we assume that the latent epigenetic age is linearly dependent on a vector of observable covariates, $X = \{X_1, \ldots, X_P\}$. We can therefore specify the following behavioral equation:

$$EA^* = X'\gamma + \omega \qquad (4)$$

where $\gamma$ is a $P \times 1$ vector of parameters and $\omega$ is an idiosyncratic error term. Plugging Equation (4) into Equation (3), we are thus able to estimate a system of $k$ reduced form regressions of the kind:

$$C_k = \lambda_k X'\gamma + \lambda_k \omega + \varepsilon_k, \qquad k = 1, \ldots, K \qquad (5)$$

Conditional on the behavioral equation being correctly specified, Structural Equation Modelling is generally more efficient because it uses all data to simultaneously estimate parameters. This simultaneous estimation minimizes potential information loss and takes into account the full covariance structure of the data.[8]

The estimation of the MEGA clock using the three methods outlined above with commonly used statistical software (e.g. Stata, R) automatically entails the standardization of the variables used to build the latent factor or, in the case of $MEGA_{SEM}$, the expression of the latent factor in the same scale of the predicted value of the first clock included in system of measurement equations.[9] However, one of the advantages of epigenetic clocks is their interpretability in years of age, a metric that is both simple and of great public-policy relevance. In order to bring the MEGA clocks back to a unit of measurement that can be interpreted in terms of years of age, we adopt a 'de-standardization' procedure, by manually computing the weighted averages in Equations (1) and (2) leaving the clocks $\{C_1, \ldots, C_K\}$ in their natural scale, years of age.

Combining information from different clocks has the advantage of reducing the measurement error associated with each individual clock, allowing us to capture the process of epigenetic aging more precisely. Minimizing measurement error and improving estimation precision is important given the limited sample sizes of most epigenetic datasets. Reducing measurement error also has the potential to mitigate the risk of attenuation bias in our estimates.

---

[8] A similar approach can be followed when the latent Epigenetic Age $EA^*$ variable is used as a control, not an outcome, in the behavioral equations. Consider the following vector of education, health, and labor-market outcomes $Y = \{Y_1, \ldots, Y_D\}$; the resulting behavioral equation would be $Y = EA^{*'}\delta + H'\theta + u$, where $H'$ is a vector of controls and $\delta, \theta$ are vectors of parameters to be estimated.

[9] Here, we use Stata v. 17, and commands "sem" for $MEGA_{SEM}$ , "factor" for $MEGA_{FA}$ and "egen weightave" for $MEGA_{WGT}$.

## 4. The Avon Longitudinal Study of Parents and Children

### 4.1. Data Overview

Our empirical analyses utilize data from the Avon Longitudinal Study of Parents and Children (ALSPAC), also known as "Children of the 90s", an English birth cohort study that provides a unique opportunity to explore the effects of genetic, environmental, and social factors on the health and development of children and their families. The study initially enrolled over 14,000 pregnant women residing in the county of Avon, U.K., with expected delivery dates between April 1991 and December 1992, resulting in 14,062 live births and 13,988 children surviving to one year of age. In the late 1990s, additional eligible mothers and children who had not joined the initial waves were recruited, bringing the total sample size to 15,447 pregnancies and 14,901 children alive at one year of age. This included 14,833 unique mothers as of September 2021, following further phases of recruitment.

ALSPAC employs a longitudinal design with multiple follow-up visits at key developmental stages. Data collection encompasses questionnaires, clinical assessments, biological sampling, and linkage to administrative records.[10] This comprehensive approach captures a wide array of information, including physical and mental health, cognitive abilities, socio-economic factors, and environmental exposures (Boyd *et al.*, 2013; Fraser *et al.*, 2013; Northstone *et al.*, 2019, 2023). The biological data collected includes genetic and epigenetic information, with blood samples taken at various time points. Longitudinal DNA methylation data for 1,022 children and their mothers have been collected using Illumina Infinium 450k arrays, the same profiling technology used in each of the clocks in the MEGA clock. For children, DNA methylation was extracted from cord blood samples at birth and from peripheral blood samples at child ages 7 and 15-19 years.[11]

---

[10] At age 18, study children were sent 'fair processing' materials describing ALSPAC's intended use of their health and administrative records and were given clear means to consent or object via a written form. Data were not extracted for participants who objected, or who were not sent fair processing materials.

[11] See Relton *et al.* (2015) for a detailed description of the Accessible Resource for Integrated Epigenomic Studies (ARIES), the sub-cohort of ALSPAC with DNA methylation data. Study data were collected and managed using REDCap (Research Electronic Data Capture) electronic data capture tools hosted at the University of Bristol (Harris

Using these data, we then compute epigenetic age from the following four epigenetic clocks: Horvath (Horvath, 2013), Hannum (Hannum *et al.*, 2013), PhenoAge (Levine *et al.*, 2018) and GrimAge (Lu *et al.*, 2019).[12] We first combine these clocks into a single measure of epigenetic age using three alternative methods (described in Section 3), and then use them separately in our three empirical applications to test the validity of our approach. Our analyses primarily rely on DNA methylation measured in late adolescence (median age 17.5), when children in our sample are biologically closer to the adult populations on which these clocks were originally developed.

## 4.2.   Estimation Samples

After conditioning on the availability of DNA methylation data and the main variables of interest (see Section 5 for more details), our final estimation samples consist of 598 observations for Application 1 (human capital); 448 observations for Application 2 (child abuse); and 597 observations for Application 3 (school-entry age). Statistics describing the socio-demographic characteristics of each estimation sample, compared to the full ALSPAC sample, are reported in Table A2. Consistent with Relton *et al.* (2015), children in the epigenetic subsamples are over selected from families with higher socio-economic status. Standard *t*-tests of differences in means between our estimation samples and the general ALSPAC sample (reported in the last three columns of Table A2) reveal that children in our estimation samples are more likely to have slightly older and more educated mothers, and fathers with a higher social class relative to the average ALSPAC child. In addition, our estimation samples are made up of a larger share of first-born children and display a small gender imbalance in Applications 1 and 2 (respectively, 59 and 62 percent of girls as compared to boys).

---

*et al.*, 2009). REDCap is a secure, web-based software platform designed to support data capture for research studies. The ALSPAC study website contains detailed information on all the available data, through a fully searchable data dictionary and variable search tool (http://www.bristol.ac.uk/alspac/researchers/our-data/). Ethical approval for the study was obtained from the ALSPAC Ethics and Law Committee and the Local Research Ethics Committees. Informed consent for the use of data collected via questionnaires and clinics was obtained from participants following the recommendations of the ALSPAC Ethics and Law Committee at the time. Consent for biological samples has been collected in accordance with the Human Tissue Act (2004).

[12] As standard in the literature, we compute GrimAge using the PCGrimAge algorithm (Higgins-Chen *et al.*, 2022).

## 5. Main Results

### 5.1. Features of the MEGA clocks

One key theoretical assumption of our approach is that a single dimension captures the common variance across existing epigenetic clocks. We can check the plausibility of this assumption leveraging the exploratory factor analysis. When performing factor analysis using the four epigenetic clocks described above (Horvath, Hannum, PhenoAge and GrimAge), only one factor appears to explain more variance in the data as compared to each single clock individually, i.e. only one factor satisfies the Kaiser criterion of having eigenvalue greater than one (Kaiser, 1960).[13] As shown in Table A3, all four clocks have factor loadings larger than 0.4 in all estimation samples (a threshold commonly used in the literature to retain items, see Stevens, 2002) and all have more than 50 percent unique variance (i.e. variance that is not explained by the remaining clocks).

The distributions of the three MEGA clocks, the four original clocks, and chronological age are displayed in Figure 1 for the estimation sample in our child abuse analysis (Application 2).[14] Descriptive statistics for the same variables can be found in Table 1, separately for the estimation samples of all empirical applications. In our analysis of child abuse, children range in age from 14.6 to 19.3 years old. The age distribution follows a bimodal distribution around ages 15 and 17 – the timing of the two clinical assessments collecting DNA methylation data in ALSPAC. The four traditional clocks which are trained on adult samples and therefore are not fine-tuned to predict adolescent age, display greater dispersion and less precise centering around the chronological age compared to the MEGA clocks. This pattern is the first rough indication that aggregation improves the signal-to-noise ratio in our epigenetic aging measures.

The predictive performance of individual clocks against chronological age, shown in Figure A2, reveals substantial heterogeneity across measures. The Hannum and Horvath clock have the highest accuracy, with linear fits that closely follow the 45-degree line representing perfect prediction. GrimAge has the largest intercept shift (GrimAge estimates of epigenetic age are on average 17.4 years higher than chronological age), but the smallest dispersion and second-closest slope to unity after Horvath.

---

[13] The selection of only one factor is unambiguous, as the second-largest factor has eigenvalue equal to 0.10.

[14] Application 2, illustrated in section 5.3, is the first in which the MEGA clocks are used as outcomes. In Figure 1, the values of $MEGA_{SEM}$ are computed as the linear prediction of the latent variable based on the coefficients from the structural model, described more in detail in section 5.3.1.

When comparing the predictive performance of the MEGA clocks against chronological age in Figure A3, all three methods exhibit a lower degree of dispersion as compared to the standard clocks and have slopes that are very close to one. Consistently, $MEGA_{SEM}$ and $MEGA_{FA}$ display the highest correlation with age in our sample, closely followed by GrimAge and $MEGA_{WGT}$ (Figure A4). All in all, the MEGA clocks appear to be better predictors of chronological age in our adolescent sample than the traditional clocks.[15] This enhanced performance reflects the variance reduction achieved through aggregation, which attenuates measurement error while preserving the signal common to all clocks.

## 5.2. Application 1: Epigenetic Aging and Human Capital

Our first empirical application speaks to the relevance of epigenetic aging in economics research, by documenting its association with measures of human capital. We show that accelerated epigenetic aging in adolescence predicts fewer years of education, worse mental health, and weaker attachment to the labor market in early adulthood – even after accounting for family background, skills, and traditional health markers. This highlights epigenetic age as a new, policy-relevant indicator of human capital formation, extending recent evidence on its predictive power for health and cognition (e.g. Faul et al., 2023; Raffington et al., 2023a).

### 5.2.1. Estimating Equation

We begin by estimating the association between accelerated epigenetic aging in late adolescence and early-adulthood cognitive and mental-health outcomes. We do so estimating the following linear regression model:

$$Y_{t+1} = \delta_0 + \delta_1 MEGA_{i,t} + \delta_2 Health_{i,t} + \theta X_i + \varepsilon_{i,t} \qquad (6)$$

where $Y_{t+1}$ is one of the four outcomes measured in early adulthood (period $t + 1$). All outcomes are coded as binary indicators for negative events, measured in the participants' mid-twenties. We consider two key human capital outcomes for young adults: (i) not having attained a university degree by age 26; and (ii) being neither in employment, education, nor training (NEET) at age 25. Our third and fourth outcomes relate to socio-emotional and mental health: (iii) an indicator for being above the diagnostic threshold of 12 for the self-assessed Short Moods and Feelings Questionnaire (SMFQ) at age 25 (Angold *et al.*, 1995), a widely used psychometric scale of mental health employed as a screening tool for depressive symptoms, with values ranging from 0 to 26; and (iv) having been diagnosed with depression by age 22. Further, $MEGA_{i,t}$ is one of the MEGA clocks for child $i$ measured in period $t$, namely late adolescence. $X_i$ is a vector of the following individual controls: mother's age at birth of the study child; binary indicators for mother's education, father's social class, and the child's gender, birth year, and birth order. Importantly, $X_i$ includes the child's age at the time of the DNA methylation assessment used to build the MEGA clocks, meaning that we can interpret the outcome as a measure of age acceleration. Lastly, $Health_{i,t}$ is a vector of individual health and behavioral outcomes measured at age 15: BMI, being a smoker and drinking.[16]

---

[15] These descriptive results presented for the estimation sample of Application 2 hold in the larger samples used in Applications 1 and 3.

A causal interpretation of $\delta_1$ requires strong assumptions: conditional on the observed covariates included in the regression, the MEGA clock is unrelated to any unobserved confounders. Although a wide set of individual and family characteristics $X_i$ are included as linear controls, other unobserved factors likely remain. Thus, the estimate of $\delta_1$ should be interpreted as a conditional association, rather than a causal effect.

In an augmented version of equation (6), we also control for two latent factors interpretable as measures of cognitive and socio-emotional skills. These are obtained by extending our Structural Equation Model with two additional sets of measurement equations, using the same procedure applied to the MEGA clock. Both factors draw on age-16 assessments. The cognitive skills factor is based on Mathematics, English, Science and aggregate test-scores, taken from linked administrative records in the National Pupil Database on General Certificate of Secondary Education (GCSE) exams, or an equivalent certification. The socio-emotional skills factor, on the other hand, is based on child self-reported SMFQ and on four subscales of the mother-reported Strengths and Difficulties Questionnaire (SDQ; Goodman *et al.*, 2000).[17]

---

[16] Being a smoker is defined as a binary indicator for smoking at least one cigarette per week, while 'Drinking' is defined as a binary indicator for any strictly positive answer to the question "Number of times young person has had 5+ full drinks in 24 hours, in the last 2 years". Both measures are reported by adolescents themselves.

### 5.2.2. Results

Estimates from equation (6) are reported in Tables 2 and 3, and illustrated in Figure 2. We find that one extra year of epigenetic age acceleration is associated with a 2 percentage points (pp) increase in the probability of not having a university degree by age 26; a 2 to 3 pp increase in the probability of being NEET at 25; a 2 pp increase in the probability of being above the diagnostic SMFQ threshold at age 25; and between a 1 to 2 pp increase in the probability of being diagnosed with depression by age 22. The observed effect sizes range from 4 to 27 percent of the mean outcome values, indicating that people with faster age acceleration in late adolescence experience a small to moderate deterioration in economic and mental-health outcomes by early adulthood. Our results on educational attainment are consistent with Mareckova *et al.* (2023), who find evidence of a modest negative association between Horvath age acceleration and IQ in a sample of young adult women.

Age acceleration is predictive of these outcomes even after controlling for negative health status and risky behaviors in adolescence (see Tables 2 and 3). The tables show estimates for coefficients $\delta_1$ and $\delta_2$ from Equation (6), both in a simplified version that does not control for adolescent health (columns 1, 3, and 5) and for the full model specification (columns 2, 4, and 6). Our result reveal that the coefficients attached to the MEGA clocks reduce only slightly after the introduction of the *Health* control vector, suggesting that epigenetic age acceleration is capturing something more than the usual health and behavioral scales that can be measured in adolescence, and provide additional predictive power in explaining early adulthood outcomes.

To benchmark the effect sizes of the associations between the MEGA clocks and early-adulthood outcomes, we include two additional sets of measurement equations to our Structural Equation Model, which are interpretable as latent factors for cognitive and socio-emotional skills, as described in section 5.2.1. Results are shown in Figure A5. We find that the MEGA clock weakly, but consistently predicts worse early-adulthood outcomes, consistent with Figure 2.[18] In contrast, the cognitive and socio-emotional factors show stronger predictive power, but mostly within their respective domains: the cognitive factor predicts better education and labor market outcomes, the socio-emotional factor better mental health. In particular, cognitive skills are strongly associated with the likelihood of completing a university degree, with a one standard-deviation increase in the cognitive skills factor linked to an 11 pp lower probability of not having a university degree by age 26. Similarly, socio-emotional skills show stronger associations with mental health indicators, with up to an 8 pp reduction in the probability of being diagnosed with depression by age 22. By comparison, the MEGA clock effects are smaller, but more consistent across the four outcomes, predicting human capital more broadly. Importantly, the associations with the MEGA clock are of comparable magnitude to those of the socio-emotional skill factor, just slightly smaller and in the opposite direction. This suggests that the MEGA clock associations, while modest, are still economically meaningful when benchmarked against established predictors of human capital development.

---

[17] The SDQ is a 25-items questionnaire developed by psychologists that is used as a screening tool for socio-emotional and behavioral problems in children and adolescents. The four subscales we use here, each measured on a 0-10 discrete scale, are those that make up the composite 'total SDQ' score: emotional symptoms, hyperactivity/inattention, peer relationship problems, and conduct problems. Higher values correspond to greater problems in each of the areas indicated by the subscale title.

## 5.3. Application 2: The Consequences of Child Abuse for Epigenetic Aging

Having established that epigenetic age predicts key early-adulthood outcomes, we next turn to its determinants, focusing on childhood shocks. Child abuse is a particularly salient event, with well-documented long-term consequences and broad epigenetic effects (Currie and Spatz Widom, 2010; Lawn *et al.*, 2018). We extend the existing literature by (i) specifically looking at epigenetic age rather than broader changes in DNA methylation, (ii) incorporating measures across different developmental periods and multiple reporters, and (iii) adding a longitudinal dimension to the analysis of child abuse. Our results show that abuse experienced before age 10 is associated with half a year of accelerated epigenetic aging in late adolescence – an effect comparable to that of growing up in socio-economic disadvantage (Fiorito *et al.*, 2017).

---

[18] The MEGA clocks and the cognitive and socio-emotional factors are standardized in the figure, to enhance comparability.

### 5.3.1.  Patterns of Child Abuse

We measure child abuse in ALSPAC using the responses to prospectively collected questions on child cruelty from mothers ($M$) and their partners ($P$).[19] Caregivers provided prospective reports of physical and emotional cruelty towards the study child, coming from themselves or their partners, several times throughout the data collection period.[20] In order to make carer-reported data more comparable to self-reported data on child abuse (described in the paragraph below), we define two developmental periods: one ranging from age 0 to 10 and the other from age 11 to 18. We then combine the carer-reported cruelty across periods and, for each rater $r$ and period $t$, with $r \in \{M, P\}$ and $t \in \{(0-10), (11-18)\}$, we define exposure to cruelty as a binary variable equal to one if rater $r$ reported any instances of child physical or emotional cruelty across the given period $t$ and zero otherwise.

Once cohort children reached adulthood (22+ years), they retrospectively reported any instances of physical or sexual abuse. These retrospective questions asked for the frequency with which adults in the family were violent towards the study child in two time windows: before the age of 11 and between the ages of 11 and 17. These questions, rated on a 5-point Likert scale from 1 'Never' to 5 'Very often', assess how frequently adults in the family were violent toward the study child. We dichotomized these responses using specific cutoff values (Table A1) and created a composite measure of self-reported cruelty for each developmental period. This binary indicator equals one if participants reported experiencing any form of physical or emotional cruelty at least once during that period, and zero otherwise. Study children were also asked about sex abuse over the same two time periods (see last two rows of Table A1 for more details on the questions). As before, we build one binary indicator of self-reported sex abuse for each developmental period, equal to one if the study child reported having experienced any sexual abuse at least once and zero otherwise.

---

[19] Similar question on whether the child was exposed to sex abuse from anyone were collected too. However, due to the sensitive nature of these questions for parents, we have opted not to include them in this analysis.

[20] Mothers and their partners were asked to report child cruelty in the periods corresponding to the following child ages: 8 months (since birth), 1.75 years (since age 8 months), 2.75 years (since age 18 months), 4 years (since age 2.5), 5 years (in past year), 6 years (since age 5), 9 years (since age 8), 9-10 years, 11 years, and 17-18 years. While we have reports from mothers in all periods mentioned above, we only observe reports of child cruelty from the mothers' partners at child ages 6 years, 9-10 years and 11 years.

Figure 3 shows the correlation between child and parent reported measures of child abuse for all ALSPAC children for whom the variables above are non-missing. While there are strong intertemporal and cross-dimensional correlations between reports from the same rater, cross-rater correlations are positive but lower in magnitude.[21] This is consistent with the literature on parental reporting versus child self-reporting of adverse childhood experiences, which highlights substantial discrepancies in reporting instances of abuse. Caregiver reports often understate the severity and frequency of adverse events compared to children's self-reports, whether the caregiver is reporting their own or their partner's behavior (Fisher *et al.*, 2011). Sibling corroboration indicates that self-reports are reliable. Newbury *et al.* (2018) and Baldwin *et al.* (2019) show that prospective parental reports and retrospective self-reports identify largely non-overlapping groups of maltreated individuals, with self-reports showing stronger associations with psychiatric problems. Using ALSPAC data, Houtepen *et al.* (2018) find that critical events like sexual abuse are underreported by parents, suggesting that children's self-reports might hold additional value. Soares *et al.* (2021) exclusively use retrospective self-reports due to the limitations of prospective parental data, while Warren *et al.* (2019) combine both, finding self-reports more significant in cases of sexual abuse. In light of the findings from the literature in developmental psychology described above, here, we adopt a comprehensive approach including both self-reports and prospective parental reports to capture the full spectrum of childhood maltreatment, accounting for potential underreporting by parents and minimizing measurement error.

Table 4 shows the prevalence of abuse in the estimation sample in two developmental periods (age 0-10 and age 11-18), as rated by the mother (header 'M'), the mother' partner ('P'), the child ('C'), and all possible combination of these three raters. Child abuse is divided between instances of child cruelty (emotional and physical), sex abuse and a combination of the two ('Any child abuse'). As we only have access to self-reported data on childhood sex abuse, in the 'Sex abuse' rows all cells but those in column (C) are empty.[22] Instances of child cruelty between child ages 0 and 10 range between 2% to 23% in the sample, depending on the rater – with children reporting higher prevalence than parents. When combining all raters together (our preferred measure), the prevalence of child cruelty increases to 33.7%. Combining this with the 3.6% self-reported cases of sex abuse leads to up to 34.8% of children in the sample having experienced some form of emotional and/or physical cruelty or sex abuse before turning 11 years old.[23] When looking at abuse during adolescence (from age 11 onwards), we find again that children report more instances of abuse as compared to their parents, with the total number of those exposed to any form of abuse according to any rater reaching almost 20 percent. There is a relatively large persistence in abuse over time: 15.6 percent of children in the sample have experienced some form of abuse in both developmental periods, suggesting that almost half of those experiencing abuse in early childhood (0-10 years old) are also victim of abuse in adolescence.

---

[21] The correlation table looks similar when restricting the sample to children with available DNA methylation information, which is how we define the estimation sample used to produce the main results of this paper (see Figure A1).

[22] 'Any child abuse' is thus computed in each column using the 'Child cruelty' measure from the (group of) rater(s) indicated in the column and child-reported sex abuse.

### 5.3.2. Estimating Equation

We study the associations between measures of child abuse over time defined above and the MEGA clocks by estimating the following linear regression model:

$$MEGA_{i,t} = \beta_0 + \beta_1 Abuse_{i,t} + \beta_2 Abuse_{i,t-1} + \gamma X_i + \varepsilon_{i,t} \qquad (7)$$

where $MEGA_{i,t}$ is again one of the MEGA clocks for child $i$ measured in period $t$, here corresponding to late adolescence. Moreover, $Abuse_{i,t}$ is a binary measure of exposure to child cruelty or child sexual abuse between child ages 11 and 18, as reported by either the mother, the mother's partner or the child herself. $Abuse_{i,t-1}$ is defined similarly for the time period going from the child's birth to age 10. Lastly, $X_i$ is the same vector of controls defined in section 5.2.1.

The previous literature on the epigenetic effects of childhood adversity leads us to expect both coefficients $\beta_1$ and $\beta_2$ to be positive. That is, we expect experiences of child cruelty or sex abuse to be associated with increased age acceleration. There is limited evidence on how abuse unfolds longitudinally across childhood, however. It is thus unclear *a priori* whether more recent abuse has a stronger association with epigenetic age acceleration in late adolescence ($\beta_1 > \beta_2$) or whether early childhood adversity correlates with a stronger, long-lasting scar ($\beta_1 < \beta_2$). Given the same logic as in our human capital analysis, here too, we interpret $\beta_1$ and $\beta_2$ as conditional associations rather than as the causal effect of child abuse on epigenetic age acceleration.

---

[23] The relatively high incidence of child abuse is potentially explained by the socio-economic gradient in parents' reports. High-SES parents, which are disproportionately represented in the ALSPAC cohort, tend to report being cruel to their children more often than low-SES parents. In order to limit confounding coming from this source, all regressions in Section 5 control for maternal education and paternal social class.

### 5.3.3. Results

Our OLS estimates of the coefficients in Equation (7) are reported in Table 5. Different versions of the MEGA clock appear across the columns: $MEGA_{SEM}$ in column 1, $MEGA_{FA}$ in column 2 and $MEGA_{WGT}$ in column 3. The results are very similar across the different versions of the MEGA clock – experiencing any form of child abuse (sexual or cruelty) between ages 0 and 10 is associated with over half a year of accelerated epigenetic aging. Abuse experienced after age 11 does not appear to be significantly associated with age acceleration – the coefficient is generally negative and small in magnitude. This negative relationship is driven by the strong intertemporal correlation of our measure of abuse. When we control for child abuse occurring in one developmental period and not the other, estimated coefficients are always positive (albeit only significant for the 0-10 age range).

The half-a-year acceleration in epigenetic aging (an increase ranging from 21 to 28 percent of a standard deviation) is within the estimated range of other studies that have looked at the association between early life adversity and age acceleration: it is comparable, for example, to the age acceleration observed for children born in low socio-economic positions (Fiorito *et al.*, 2017). Marini *et al.* (2020) estimate 1 to 2 months of accelerated aging at age 7 for ALSPAC children exposed to sexual and physical abuse, a number that our estimates suggest might compound over time. This compounding effect is consistent with the effect sizes found Lawn *et al.* (2018), who show that childhood exposure to sexual and physical abuse correlates with 2.7 to 3.4 higher epigenetic age as adults (29 to 47 years old).

When we disaggregate child abuse into child cruelty and sex abuse, we find that only child cruelty between ages 0 and 10 is strongly and precisely associated with epigenetic age acceleration (see Table 6). The positive coefficient amounts to about half a year greater age acceleration for children exposed to child cruelty by age 11, similar in magnitude to the effects reported in Table 5. Estimated coefficients for sex abuse are also positive for both periods, albeit about half the size and imprecisely estimated (statistically insignificant at conventional thresholds). This suggests that the epigenetic trace left by the combined abuse measure may be driven by child cruelty more than sex abuse. This interpretation requires caution, however, given the composition of our abuse measure. As child cruelty represents the largest component of overall abuse (as detailed in Table 4) and only 3.6 percent of the sample reports having experienced sexual abuse between ages 0 and 10, our analyses may lack sufficient statistical power to detect significant effects of sex abuse in isolation.

## 5.4. Application 3: The Effect of School-Entry Age on Epigenetic Aging

We explore the determinants of epigenetic age acceleration in our third, and final, empirical application. Specifically, we exploit the U.K. school-entry age cutoff in a regression discontinuity design to isolate the causal impact of delayed school entry on epigenetic age acceleration at age seven. This approach, standard in economics for studying the effects of school starting age on cognitive and labor market outcomes (e.g., Bedard and Dhuey, 2006; Black, Devereux, and Salvanes, 2011; Fredriksson and Öckert, 2014), allows us to provide the first causal estimates of the impact of the timing of school entry on biological development. We are also one of the first to examine the causal effects of quasi-exogenous shocks or policy changes on epigenetic aging (see Schmitz and Duque, 2022, for a related exception). We find that delayed school entry leads to accelerated epigenetic aging for children from low-SES backgrounds and small anti-aging effect for children from high-SES families.

### 5.4.1. Estimating Equation

We use a sharp regression discontinuity design (RDD) based on the school-entry cutoff in the U.K. to estimate the causal effect of delayed school entry on epigenetic age acceleration, Children born on or after September 1st are required to enter school a year later than those born just before this cutoff. This rule induces quasi-random variation in school starting age among children born within a narrow window around the cutoff, which we exploit for identification. Our estimating equation is the following:

$$MEGA_{i,t} = \theta_0 + \theta_1 Treat_i + \theta_2 MoB_i + \theta_3 Treat_i \times MoB_i + \mu W_i + v_{i,t} \qquad (8)$$

where $Treat_i$ is a binary indicator for being born on or after September 1st of a given year and $MoB_i$ is the running variable, i.e. the child's month of birth normalized to 0 in September (thus taking values -1 for August, 1 for October, etc.). As previously, $MEGA_{i,t}$ is child $i$'s MEGA clock measured at time $t$, which in this case we take as age 7 – the earliest DNA methylation measurement point since school entry. Lastly, $W_i$ is a vector of exogenous child controls, namely the child's gender, age at the time of DNA methylation measurement and their year of birth. In our main specification, we restrict the analysis to children born within four months around the cutoff, that is those born from May to December. To test for the robustness of our results, we additionally estimate equation (8) using children born within three and within two months from the cutoff.

The parameter $\theta_1$ captures the effect of being born after the cutoff (and thus entering school a year later) on epigenetic aging, measured by the MEGA clock at age 7. While we may observe a non-zero average treatment effect, it is also plausible that the impact of delayed school entry varies according to children's home or childcare environments. Specifically, children from high socio-economic status (SES) backgrounds may benefit from delayed entry, whereas those from lower SES backgrounds might experience more favorable developmental conditions within the school setting than at home (Anderson *et al.*, 2011; Holford and Rabe, 2022). To account for potential treatment effect heterogeneity by SES, we also run an augmented model specification controlling for an occupation-based measure of paternal social class at the time of the child's birth and interacting it with $Treat_i$, $MoB_i$ and their interaction.

### 5.4.2. Results

We first look at the distribution of age acceleration by month of birth in the estimation sample. We compute age acceleration as the residuals from regressions of the MEGA clock on chronological age, run separately for each of the three time points at which DNA methylation is observed (child's birth, age 7 and age 15-19). For simplicity, Figure A6 only shows results for the MEGA factor clock. As expected, there is no discontinuity in age acceleration by birth month at the time of birth, a result that primarily serves as a placebo test.[24] In contrast, we observe clear discontinuities at the school-entry cutoff when age acceleration is measured at ages 7 and 15, with higher values for children born in September compared to those born in August. These patterns provide a first piece of evidence consistent with an effect of school-entry age on epigenetic aging.

Moving to a regression framework, Panel A of Table 7 shows results from estimating equation (8). Consistent with the evidence from Figure A6, our RDD estimates indicate that delaying school entry by one year leads to a statistically significant acceleration in epigenetic age of 0.3 to 0.6 years, measured at age 7.

---

[24] Because the epigenetic clocks used to build the MEGA clock (Hannum, Horvath, PhenoAge, and GrimAge) were primarily developed and validated in adult populations, their application in children – particularly in newborns – should be interpreted with caution.

We disaggregate the main results by paternal social class at birth (results using the MEGA factor clock are shown in the blue estimates in Figure 5) and find that the average detrimental effect of delaying school entry on epigenetic aging is driven entirely by children from low-SES backgrounds. No significant impact is observed among children from higher-SES families. These findings suggest that earlier access to structured school environments may offer protective biological effects, especially for socio-economically disadvantaged children.

This study provides the first causal evidence linking formal schooling to biological aging processes during early childhood. These findings highlight the importance of early institutional exposure not only for cognitive and social development, as previously documented in the economics literature (e.g., Bedard and Dhuey, 2006; Black, Devereux, and Salvanes, 2011; Fredriksson and Öckert, 2014), but also for objectively measurable biological outcomes.

## 5.5. Sensitivity Tests

We first compare results from our three empirical applications with those that would be obtained by using traditional epigenetic clocks individually, rather than the MEGA clock. Figure A7 presents regression coefficients for the four clocks on early-adulthood human capital outcomes (Application 1). The results indicate that the patterns shown in Figure 2 are driven by the GrimAge clock, which is the only one to yield point estimates significantly larger than zero at least at the 10 percent level. In comparison to models based on individual clocks, regressions using the MEGA clock yield more precisely estimated and larger coefficients, consistent with a reduction in classical measurement error.

We turn now to consider our analysis of child abuse (Application 2). Estimates of coefficients $\beta_1$ and $\beta_2$ from Equation (7) for each of the outcomes indicated in the legend are reported in Figure 4. First, we present results for each of the single clocks used to build the MEGA clock (blue round markers) and, second, we replicate the estimated coefficients for our MEGA clocks (see Table 5) for convenience (red diamond markers). Estimates of child abuse before age 11 using traditional clocks result in larger average magnitudes than those estimated with the MEGA clocks, with point estimates ranging from 0.52 to 1.00. The MEGA clocks produce results that are the closest to GrimAge (the one they correlate with the most, as shown in Figure A4), but have smaller standard errors on average, suggesting that – all else equal – our aggregating procedures do indeed reduce measurement error as compared to using any one of the single clocks separately. In Figure A8, we replicate the same exercise but disaggregate child abuse into its two components: cruelty and sex abuse (similar to Table 6). Results are again consistent with those from Figure 4. Estimates based on the MEGA clocks have smaller confidence intervals on average than the traditional clocks, while point estimates are aligned with those from the GrimAge clock.

Lastly, we repeat this exercise for our analysis of school-entry age (Application 3), replicating results from Panel A of Table 7 (see Figure A9). The estimated treatment effects of being born after August 31$^{st}$ are positive for all individual clocks except Hannum's, with point estimates from the MEGA clocks falling within the range of the individual clocks' ones. Here again we confirm that using the MEGA clocks yields smaller standard errors than each of the four traditional clocks individually, thereby improving precision.

To test whether results are sensitive to the exclusion of either one clock used in the MEGA algorithms, we compute leave-one-out versions of the MEGA clocks. Exploratory factor analysis confirms a uni-factor model for any combination of three out of the four clocks in all estimation samples. Results for our analysis of human capital are displayed in Figure A10. The point estimates from even-numbered columns of Tables 2 and 3 remain quite robust when excluding either one of the Hannum, Horvath or PhenoAge clocks. Interestingly, our point estimates converge to zero when excluding the GrimAge clock from the MEGA, indicating that most of the associations between epigenetic age acceleration and early-adulthood outcomes that we observe run through this clock. Results for our analysis of child abuse are also robust to the exclusion of either one clock (see Figure A11). As expected, this comes at the expenses of precision, the loss of which is greater when excluding the GrimAge clock. All point estimates for early-childhood abuse are remarkably stable and they remain positive and statistically different from zero at least at the 10 percent level. We also apply this leave-one-out computational strategy to our analysis of the effects of school-entry age (see Figure A12). Results are presented only for the factor analysis and weighted index approach, due to a lack in convergence for the Structural Equation Models. Here too results appear to be remarkably robust to the exclusion of either one clock, remaining close to the treatment effects displayed in Panel A of Table 7. The only exception appears to be Horvath's clock, which is also the one displaying the largest point estimates in Figure A9.

We then turn to consider the sensitivity of our results to our measurement of exposure to child abuse. As argued in section 4.1, our preferred measure of abuse includes information from all available raters, namely mothers, their partners and the child herself. Table A4 shows that our results in Table 5 are not driven by this measurement choice: the estimated coefficients attached to abuse between ages 0 and 10 are stable across raters and combination of raters, roughly ranging from 0.3 to 0.5. The use of reports from all raters (last column of Table A4) results in the lowest standard errors across all methods, suggesting that harnessing these different sources of information can help reduce the measurement error linked to the under-reporting of sensitive constructs.

Next, we examine the sensitivity of our school-entry age results to bandwidth selection. Theoretically, smaller bandwidths imply lower bias in the causal estimates but bigger confidence intervals, due to the smaller sample size. Restricting the bandwidth from four to three months around the cutoff (Panel B of Table 7) yields similar point estimates as the main results in Panel A, although they are no longer statistically different from zero due to the loss in precision. Further narrowing the sample to children born between July and October (Panel C of Table 7) results in a doubling of the standard errors but also of the point estimates; focusing only on those born just before or just after the cutoff – the cleanest approach from a theoretical perspective – suggests that delaying school entry increases epigenetic aging by 0.9 to 1.2 years.

Finally, we assess whether the aging penalty associated with delayed school entry persists beyond age 7, into adolescence. Table A5 indicates that it does not: while point estimates remain positive on average, they are much smaller (one third to one fifth of the results at age 7) and noisily estimated, such that none of them is statistically different from zero at conventional levels. The epigenetic aging differences induced by the school-entry cutoff fade over time, mirroring a common finding in the literature that the effects of school starting age on various outcomes tend to weaken as children age (e.g. Bedard and Dhuey, 2006).

## 6. Potential Mechanisms

We now discuss how epigenetic aging may serve as a biological pathway connecting early-life exposures, including child abuse and age at school-entry, to downstream outcomes in cognition and mental health. Due to the limited range of biomarkers in our data, we are constrained in our ability to directly investigate biological channels beyond DNA methylation. Consequently, we draw on insights from the molecular epigenetics literature and observed behavioral patterns to provide a coherent narrative of how environmental exposures may translate into accelerated epigenetic aging and later into worse adult outcomes. The goal is not to identify definitive causal pathways, but to offer biologically and behaviorally grounded interpretations of our findings.

### 6.1. Human Capital

Accelerated epigenetic aging is likely to influence later cognitive and mental-health outcomes through multiple interconnected biological pathways. The machine learning approaches used to develop epigenetic clocks are fundamentally agnostic to the underlying biology. This means that the CpG sites they rely on capture a wide range of processes rather than a single mechanism. These sites regulate cell growth and survival, inflammation, antiviral responses, and DNA repair (Horvath, 2013; Hannum *et al.*, 2013; Levine *et al.*, 2018), and are over-represented in gene sets involved in the immune system, lipid function, and adipocytes communication (Lu *et al.*, 2019). Although the adults in our sample are still too young to show clinical aging traits, the molecular processes captured by epigenetic clocks could already affect brain function and correlate with early cognitive deterioration (Felt *et al.*, 2023). Some of the most central gene regions in epigenetic clocks are involved in neuronal pathways, including neurogenesis, neuron differentiation, and neuron death (Han *et al.*, 2018), providing a potential biological link to learning, memory, and other human-capital outcomes.

### 6.2. Child Abuse

Child abuse may accelerate epigenetic aging through both biological and behavioral stress responses. One plausible biological mechanism involves changes in the proportion of immune cells, particularly leukocytes, which in turn affects methylation patterns (Lima *et al.*, 2022).[25]

---

[25] Since distinct blood cell types exhibit unique methylation patterns, cell counts are typically controlled for in epigenetic analyses. Yet, these counts may function as 'bad controls', as they mediate the relationship between environmental exposures and the epigenome.

Child abuse, in particular, has been consistently associated with changes in immune cell proportions, driven by heightened inflammatory activity (D'Elia *et al.*, 2018; Renna *et al.*, 2021). To test whether blood cell counts mediate the relationship between child abuse and epigenetic aging in our sample, we compute standard cell counts for peripheral blood samples from the age 15-19 DNA methylation data in ALSPAC,[26] using the Houseman *et al.* (2012) method. In our estimation sample, children exposed to abuse at least once display substantially larger associations between blood cell counts and age acceleration (Figure A13). When controlling for blood cell counts in our main regressions, the magnitude of the estimated coefficients decreases by approximately one-third, suggesting a mediating role (Table A6).

Beyond cellular changes, stress-related pathways involving the hypothalamic-pituitary-adrenal axis and cortisol production are also likely involved (Dammering *et al.*, 2021; Suarez *et al.*, 2018). Individual differences in resilience – through emotion regulation and self-control – have also been shown to moderate stress induced epigenetic aging (Harvanek *et al.*, 2021, 2023).[27] Risky behaviors and lifestyle changes can also play a role in how stress affects epigenetic aging (Schmitz *et al.*, 2022; Jung *et al.*, 2023). Both biological and behavioral stress responses can additionally affect other hallmarks of aging, including circadian rhythms, immune system functioning, and nutrition, which subsequently influence the epigenome (Harvanek *et al.*, 2023).

### 6.3. School-Entry Age

We show that structured educational environments appear to influence epigenetic aging, particularly among socioeconomically disadvantaged children. While much of the existing research has focused on non-cognitive skills, academic performance, and behavioral outcomes (e.g., Bedard and Dhuey, 2006; Fredriksson and Öckert, 2014), systematic evidence on the potential protective health effects of early exposure to structured educational environments remains limited. Some studies, such as Anderson *et al.* (2011), suggest that the transition from a relatively unstructured home environment to the routine and regulation of school life may reduce snacking opportunities and promote physical activity – plausibly contributing to improved physical health. Holford and Rabe (2022) similarly find that children assessed later in the school year (and thus exposed to more schooling) display lower weight-for-height ratios, suggesting cumulative effects of structured schooling on health-related behaviors.

---

[26] The cell types considered here are B-lymphocytes (Bcell), CD4+ T-lymphocytes (CD4T), CD8+ T-lymphocytes (CD8T), granulocytes (Gran), monocytes (Mono), and natural killers (NK).

[27] Research on posttraumatic stress disorder (PTSD) suggests that specific symptom clusters, such as emotional withdrawal, sleep problems, and cognitive dysfunction, correlate more strongly with changes in epigenetic aging than the overall PTSD severity (Katrinli *et al.*, 2020; Na *et al.*, 2022).

Building on these intuitions, we further examine treatment effect heterogeneity by SES in our data. To test whether school systematically offers healthier environments for low-SES children, we analyze a range of health and nutrition outcomes in the full ALSPAC sample in Figure 5. Children from manual occupation households who enter school later because of their birthdate display not only accelerated epigenetic aging, but also worse general health (as reported by their mothers), and higher consumption of processed foods and sugary and fatty items at age 7.[28] Conversely, children from higher social classes benefit from delayed school entry, exhibiting healthier dietary patterns. While all effects point in the same direction, the magnitude of the impact on epigenetic aging is substantially larger, suggesting that it may capture broader or more cumulative biological consequences than these behavioral outcomes – highlighting the value of measuring it directly. These results suggest that structured school environments may serve as a compensatory setting protecting the health and epigenetic aging trajectories of disadvantaged children, supporting the hypothesis that early-life social environments can mitigate the biological embedding of inequality.

## 7. Conclusion

As economists work to develop richer models of human behavior, there is growing interest in measuring and analyzing the epigenetic processes through which people's environments affect their biological functioning. A focus on biological, rather than chronological, age opens the door to a deeper understanding of people's health and disease risk, physical functioning, and cognitive performance as they age. Epigenetic clocks have emerged as the leading tool for summarizing age-related epigenetic markers – now available in many standard large-scale, population-representative data sets – into a single measure of biological age.

---

[28] In Figure 5, 'high-sugar, high-fat diet' refers to a standardized dietary pattern score derived by Ambrosini *et al.* (2016) using reduced rank regression, associated with more energy density, higher percent energy from free sugars and total fat, and lower density of fiber. Similarly, 'Processed diet' is a dietary pattern score indicating high consumption of processed foods, chips and soft drinks, as derived by Smith *et al.* (2013).

We make an important methodological contribution by developing a new metric – the MEGA clock – which allows researchers to combine established epigenetic clocks to increase the estimation precision in models of the determinants and consequences of epigenetic aging. Importantly, the methodological approach we propose is flexible and designed to accommodate future epigenetic clocks as they are developed. The results of our empirical applications not only establish the validity and robustness of the MEGA clock, but also the usefulness of epigenetic clocks in understanding the biological mechanisms through which socio-economic factors influence health and human capital.

The results of our empirical applications add to the growing evidence that epigenetic age is a policy-relevant measure of health and well-being. We are led to two key conclusions. First, there is a potential for epigenetics to be one mechanism linking environmental conditions in one lifecycle stage to social and economic well-being in the next. For example, we find that early-life exposure to child abuse is associated with accelerated adolescent epigenetic aging, highlighting the biological impact of early-life adversity. Accelerated epigenetic aging in adolescence, in turn, predicts worse cognitive and mental-health outcomes in early adulthood with lifetime implications for health and productivity.

Second, although many of the empirical relations we examine are correlational, we are among the first to provide some causal evidence of the effect of childhood events on the pace of epigenetic aging. Our research breaks new ground by using a regression discontinuity design to identify the causal impact of the timing of school entry on epigenetic aging. Children starting school one year later experience faster epigenetic aging, particularly if they are from low-SES backgrounds. Early access to structured educational environments may be a protective factor against the biological embedding of stressful or less structured home settings.

Our exploration of potential mechanisms suggests that accelerated epigenetic aging likely reflects both biological and behavioral responses to early-life experiences. For instance, exposure to child abuse appears to alter immune cell profiles and stress-response systems, while early school entry may provide structured environments that promote healthier routines, particularly for disadvantaged children.

More broadly, our findings underscore the potential for epigenetic measures to reshape how economists study health, well-being, and human capital development. By tracing how early-life environments become biologically embedded, epigenetic clocks offer an avenue to evaluate not only long-run consequences of adversity but also the protective effects of early interventions. The MEGA clock offers an easily interpretable, reliable tool for studying the link between environmental factors and biological aging processes, without requiring specialized training in molecular biology, enabling more targeted and effective policy interventions. Future research should continue refining these measures and exploring their applications, investigating persistence across the life cycle, testing the generalizability of results across populations, and linking specific policies – such as preschool or nutrition programs – to biological aging trajectories. By bridging biology and economics, our work contributes to laying the groundwork for a more comprehensive understanding of human development and the complex interplay between nature and nurture.

# Figures and Tables

Figure 1: Distribution of epigenetic age by clock and chronological age

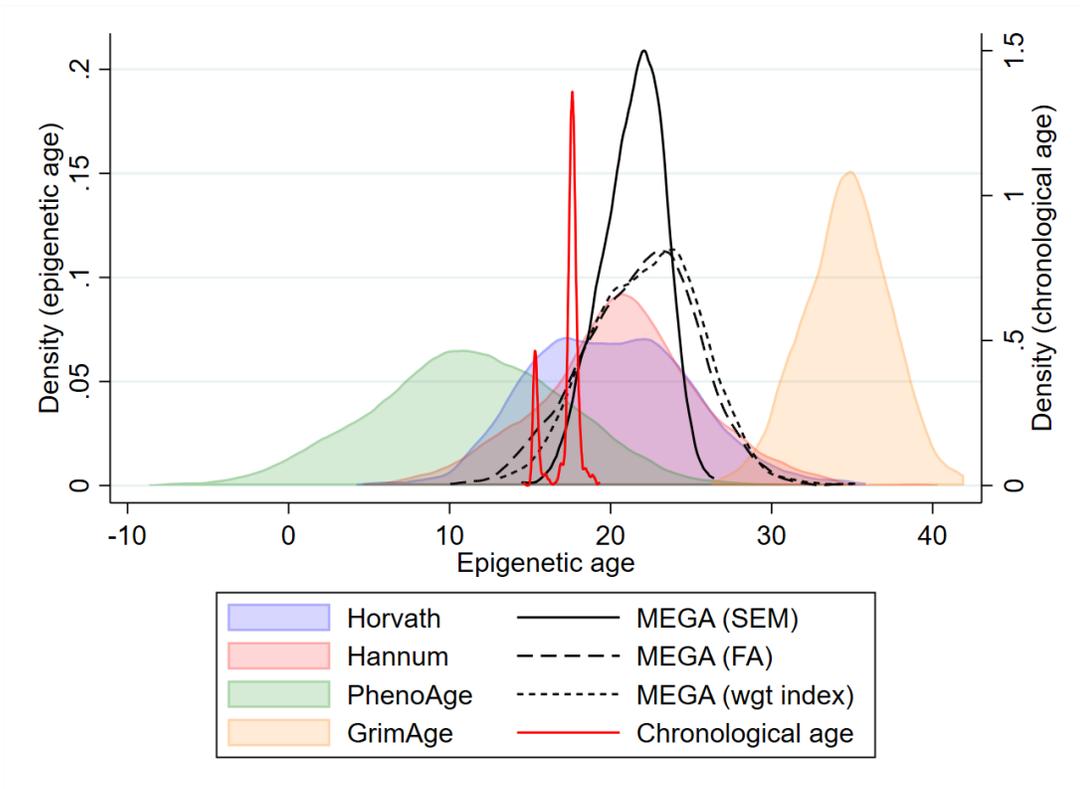

*Notes:* The figure displays the distribution of epigenetic age as measured by four individual clocks (Horvath, Hannum, PhenoAge, and GrimAge), as well as the three MEGA composite measures: MEGA (SEM), MEGA (FA), and MEGA (weighted index) constructed for the estimation sample in Application 2 (N=448). The red line represents the distribution of chronological age in the sample (measured in two waves around age 15–19).

Figure 2: Age acceleration and early-adulthood outcomes

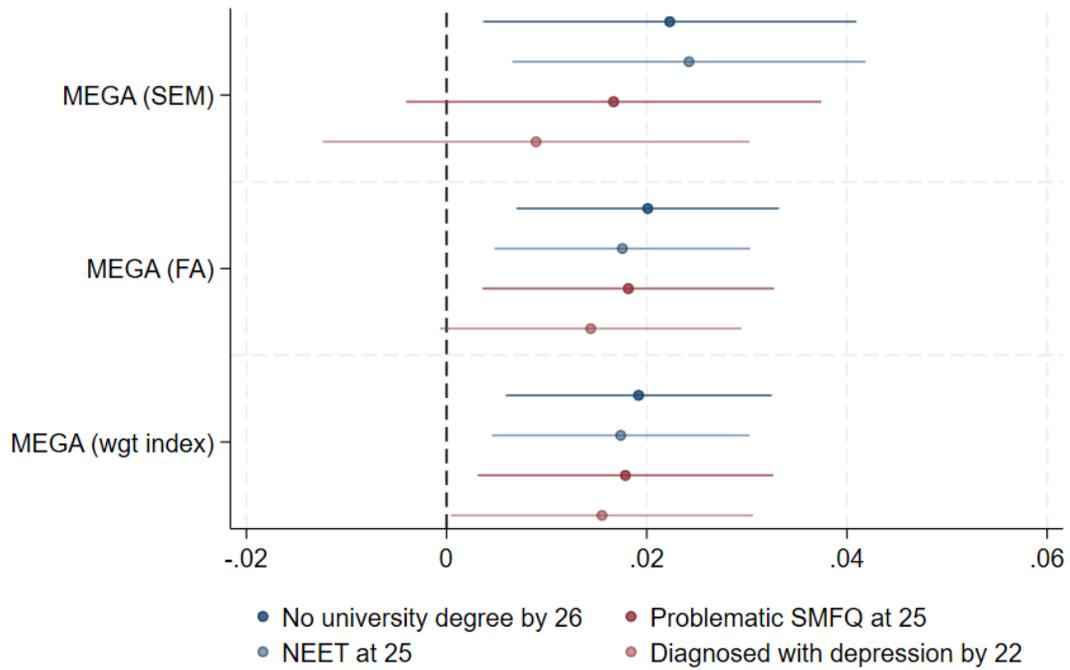

*Notes:* The figure displays point estimates for the MEGA clocks from columns (2), (4), and (6) of Tables 2 and 3. All regressions control for mother's age at birth of the study child and binary indicators for mother's education, father's social class, and the child's gender, age, birth year, and birth order. Health outcomes (BMI, smoking and drinking) are additionally controlled for. Horizontal spikes are for 90 percent confidence intervals.

Figure 3: Cross-rater, intertemporal correlations of child cruelty and sex abuse

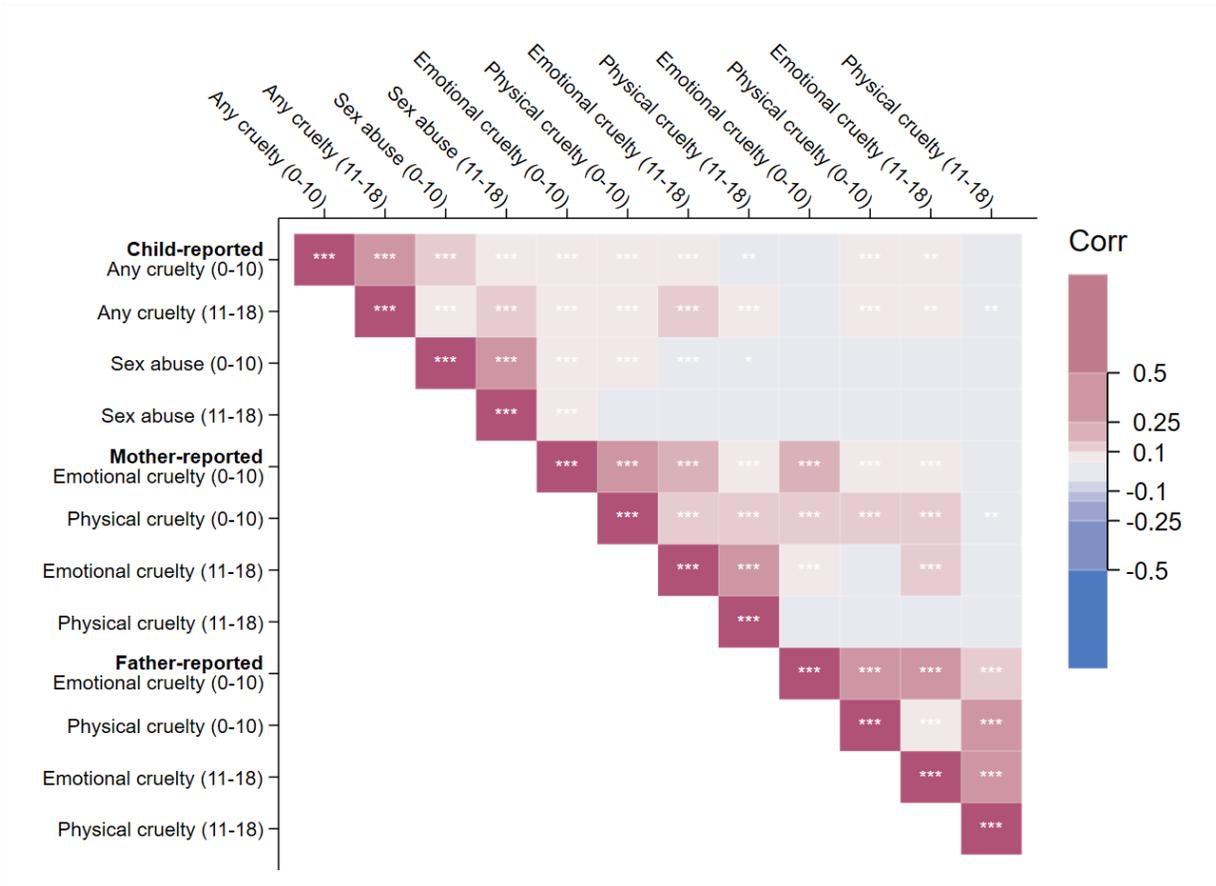

*Notes:* The correlation matrix is based on the sample of 3937 children in ALSPAC for whom all measures above are available.

Figure 4: Child abuse and age acceleration: individual clocks

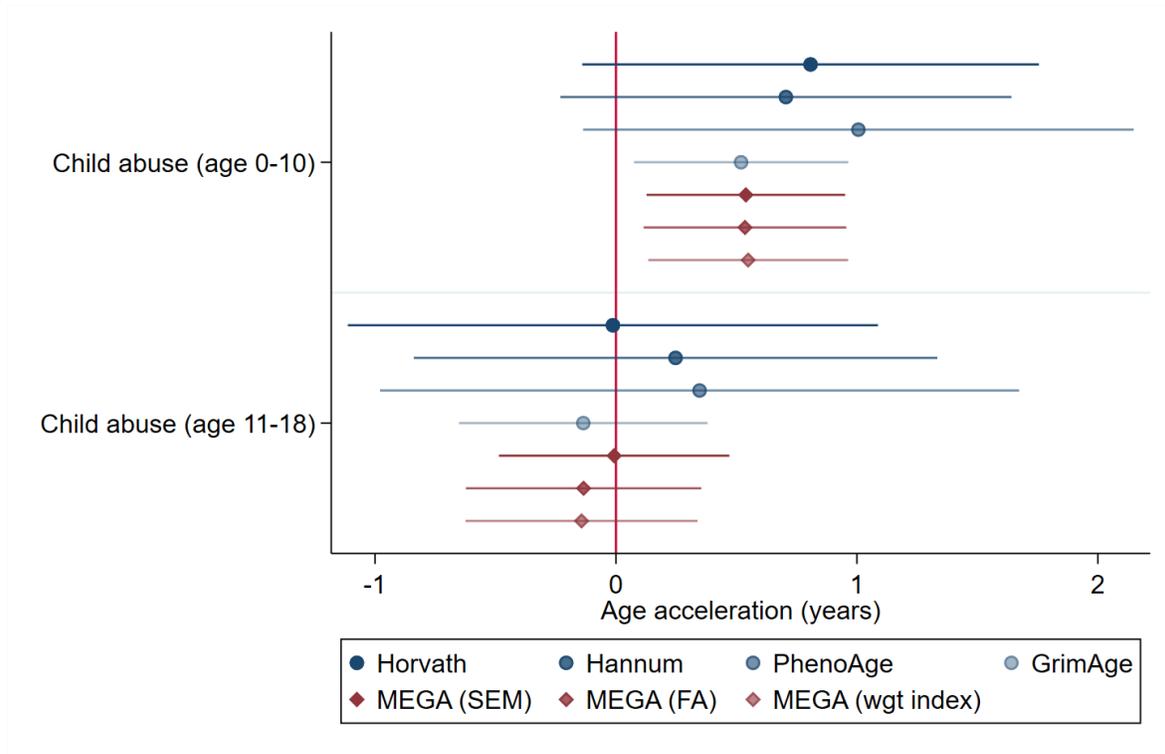

*Notes:* The blue dots in the figure replicate point estimates from Table 5 using the individual clocks as dependent variables, while the red dots report results for the MEGA clocks. All regressions control for mother's age at birth of the study child and binary indicators for mother's education, father's social class, and the child's gender, age, birth year, and birth order. Horizontal spikes are for 90 percent confidence intervals.

Figure 5: The effect of delayed school entry on age-7 outcomes, by paternal social class

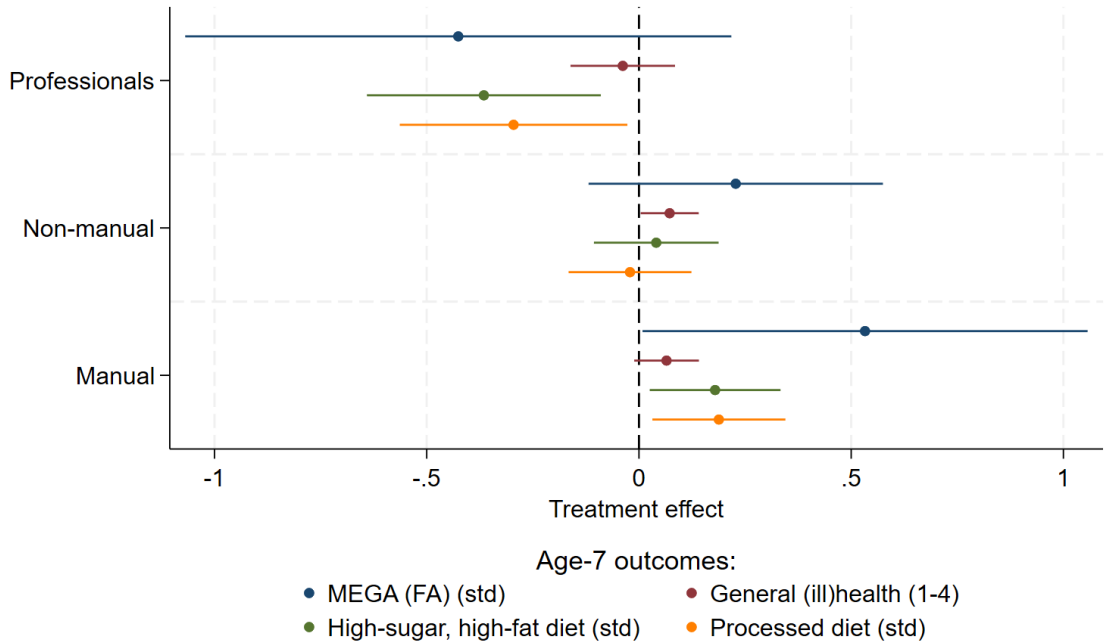

*Notes:* The figure displays estimated treatment effects by paternal social class, for all outcomes displayed in the legend. Estimates come from a model specification of equation (8) in which paternal social class is controlled for and where the treatment status, the month of birth and their interaction is additionally interacted for paternal social class dummies. All regressions control for the child's age and dummies for the child's gender and birth year. Horizontal spikes are for 90 percent confidence intervals.

Table 1: Descriptive statistics of chronological age and epigenetic age

| | $N$ | Mean | Std. Dev. | Min | Max |
|---|---|---|---|---|---|
| **Panel A: Application 1** | | | | | |
| Chronological age | 598 | 17.18 | 1.00 | 14.6 | 19.3 |
| Horvath | 598 | 19.9 | 4.92 | 4.2 | 35.8 |
| Hannum | 598 | 20.47 | 5.00 | 4.6 | 40.3 |
| PhenoAge | 598 | 11.05 | 6.23 | -11.4 | 34.7 |
| GrimAge | 598 | 34.53 | 2.73 | 26.1 | 43.5 |
| MEGA clocks | | | | | |
| $MEGA_{SEM}$ (GrimAge) | 598 | 0 | 1.85 | -6.5 | 5.5 |
| $MEGA_{SEM}$ (Horvath) | 598 | 0 | 1.51 | -5.3 | 4.5 |
| $MEGA_{SEM}$ (Hannum) | 598 | 0 | 2.97 | -10.4 | 8.8 |
| $MEGA_{SEM}$ (PhenoAge) | 598 | 0 | 3.21 | -11.3 | 9.5 |
| $MEGA_{FA}$ | 598 | 33.09 | 2.56 | 25.6 | 41.2 |
| $MEGA_{WGT}$ | 598 | 32.92 | 2.52 | 26 | 40.7 |
| **Panel B: Application 2** | | | | | |
| Chronological age | 448 | 17.23 | 0.95 | 14.6 | 19.3 |
| Horvath | 448 | 19.97 | 4.95 | 4.2 | 35.8 |
| Hannum | 448 | 20.62 | 5.03 | 4.6 | 40.3 |
| PhenoAge | 448 | 11.29 | 6.15 | -8.7 | 34.7 |
| GrimAge | 448 | 34.64 | 2.67 | 26.3 | 42.0 |
| MEGA clocks | | | | | |
| $MEGA_{SEM}$ (GrimAge) | 448 | 21.40 | 1.90 | 14.5 | 26.4 |
| $MEGA_{SEM}$ (Horvath) | 448 | 16.45 | 1.46 | 11.1 | 20.3 |
| $MEGA_{SEM}$ (Hannum) | 448 | 33.27 | 2.95 | 22.5 | 41.1 |
| $MEGA_{SEM}$ (PhenoAge) | 448 | 34.54 | 3.07 | 23.3 | 42.7 |
| $MEGA_{FA}$ | 448 | 33.78 | 2.55 | 26.3 | 40.8 |
| $MEGA_{WGT}$ | 448 | 33.19 | 2.50 | 26.6 | 40.6 |
| **Panel C: Application 3** | | | | | |
| Chronological age | 597 | 7.46 | 0.14 | 7.1 | 8.8 |
| Horvath | 597 | 11.03 | 3.42 | 2.6 | 30.8 |
| Hannum | 597 | 9.49 | 4.54 | -3.6 | 32.8 |
| PhenoAge | 597 | -1.34 | 6.74 | -23 | 20.8 |
| GrimAge | 597 | 26.1 | 2.48 | 19.6 | 33.2 |
| MEGA clocks | | | | | |
| $MEGA_{SEM}$ (GrimAge) | 597 | 1.86 | 1.07 | -1.5 | 5.6 |
| $MEGA_{SEM}$ (Horvath) | 597 | 9.4 | 1.59 | 4.3 | 14.6 |
| $MEGA_{SEM}$ (Hannum) | 597 | 12.92 | 2.18 | 5.9 | 20 |
| $MEGA_{SEM}$ (PhenoAge) | 597 | 19.41 | 3.28 | 8.8 | 30.1 |
| $MEGA_{FA}$ | 597 | 21.96 | 2.05 | 15.7 | 30.1 |
| $MEGA_{WGT}$ | 597 | 22.9 | 2.02 | 16.3 | 31.6 |

*Notes:* Descriptive statistics refer to the largest estimation sample for each of the three applications. For the second and third applications, the mean values of the $MEGA_{SEM}$ are computed as the linear prediction of the latent variable based on the coefficients from the structural model. In the first application, the $MEGA_{SEM}$ factors are instead centered on zero, as they appear on the right-hand side of the structural equation and no linear prediction can be computed for them.

Table 2: Age acceleration and early-adulthood cognitive outcomes

| | SEM | | FA | | Wgt Index | |
|---|---|---|---|---|---|---|
| | (1) | (2) | (3) | (4) | (5) | (6) |
| **Panel A. No University degree by 26** | | | | | | |
| MEGA | 0.023** | 0.022** | 0.023*** | 0.020** | 0.022*** | 0.019** |
| | (0.011) | (0.011) | (0.008) | (0.008) | (0.008) | (0.008) |
| | | | | | | |
| BMI at age 15 | | -0.001 | | -0.000 | | -0.000 |
| | | (0.006) | | (0.006) | | (0.006) |
| | | | | | | |
| Smoking at age 15 | | 0.240*** | | 0.233*** | | 0.231*** |
| | | (0.073) | | (0.074) | | (0.074) |
| | | | | | | |
| Drinking at age 15 | | -0.007 | | -0.006 | | -0.006 |
| | | (0.037) | | (0.038) | | (0.038) |
| Observations | 525 | 525 | 525 | 525 | 525 | 525 |
| Adjusted R-squared | | | 0.148 | 0.166 | 0.147 | 0.165 |
| **Panel B. NEET at 25** | | | | | | |
| MEGA | 0.028** | 0.024** | 0.019** | 0.018** | 0.018** | 0.017** |
| | (0.011) | (0.011) | (0.008) | (0.008) | (0.008) | (0.008) |
| | | | | | | |
| BMI at age 15 | | 0.006 | | 0.007 | | 0.007 |
| | | (0.005) | | (0.005) | | (0.005) |
| | | | | | | |
| Smoking at age 15 | | 0.003 | | 0.002 | | -0.001 |
| | | (0.075) | | (0.077) | | (0.077) |
| | | | | | | |
| Drinking at age 15 | | -0.047 | | -0.046 | | -0.046 |
| | | (0.036) | | (0.037) | | (0.037) |
| Observations | 352 | 352 | 352 | 352 | 352 | 352 |
| Adjusted R-squared | | | 0.012 | 0.035 | 0.011 | 0.035 |

*Notes:* Standard errors in parentheses. 'SEM' stands for Structural Equation Modelling, 'FA' for Factor Analysis, and 'wgt index' for weighted index. All regressions control for mother's age at birth of the study child and binary indicators for mother's education, father's social class, and the child's gender, age, birth year, and birth order. Mean outcome values in the estimation samples are 0.24 for not having a university degree and 0.11 for being NEET. $^{*} p < 0.1$, $^{**} p < 0.05$, $^{***} p < 0.01$.

Table 3: Age acceleration and early-adulthood mental health outcomes

|  | SEM | | FA | | Wgt Index | |
|---|---|---|---|---|---|---|
|  | (1) | (2) | (3) | (4) | (5) | (6) |
| **Panel A. Problematic SMFQ at 25** | | | | | | |
| MEGA | 0.015 | 0.017 | 0.017* | 0.018** | 0.016* | 0.018** |
|  | (0.013) | (0.013) | (0.009) | (0.009) | (0.009) | (0.009) |
| BMI at age 15 |  | -0.002 |  | -0.002 |  | -0.002 |
|  |  | (0.006) |  | (0.006) |  | (0.006) |
| Smoking at age 15 |  | 0.075 |  | 0.065 |  | 0.062 |
|  |  | (0.082) |  | (0.084) |  | (0.084) |
| Drinking at age 15 |  | -0.063 |  | -0.063 |  | -0.063 |
|  |  | (0.041) |  | (0.041) |  | (0.041) |
| Observations | 434 | 434 | 434 | 434 | 434 | 434 |
| Adjusted R-squared |  |  | 0.021 | 0.027 | 0.021 | 0.026 |
| **Panel B. Diagnosed with depression by 22** | | | | | | |
| MEGA | 0.010 | 0.009 | 0.015* | 0.014 | 0.016* | 0.016* |
|  | (0.013) | (0.013) | (0.009) | (0.009) | (0.009) | (0.009) |
| BMI at age 15 |  | 0.006 |  | 0.005 |  | 0.005 |
|  |  | (0.007) |  | (0.007) |  | (0.007) |
| Smoking at age 15 |  | 0.032 |  | 0.022 |  | 0.020 |
|  |  | (0.083) |  | (0.085) |  | (0.085) |
| Drinking at age 15 |  | 0.043 |  | 0.043 |  | 0.043 |
|  |  | (0.041) |  | (0.042) |  | (0.042) |
| Observations | 451 | 451 | 451 | 451 | 451 | 451 |
| Adjusted R-squared |  |  | 0.000 | 0.007 | 0.000 | 0.008 |

*Notes:* Standard errors in parentheses. 'SEM' stands for Structural Equation Modelling, 'FA' for Factor Analysis, and 'wgt index' for weighted index. All regressions control for mother's age at birth of the study child and binary indicators for mother's education, father's social class, and the child's gender, age, birth year, and birth order. Mean outcome values in the estimation samples are 0.18 for SMFQ and 0.20 for depression. $^*$ $p < 0.1$, $^{**}$ $p < 0.05$, $^{***}$ $p < 0.01$.

Table 4: Prevalence of abuse in the Application 2 estimation sample

| | M (1) | P (2) | C (3) | MP (4) | CM (5) | CP (6) | CMP (7) |
|---|---|---|---|---|---|---|---|
| **Age 0-10** | | | | | | | |
| Child cruelty | 13.8% | 2.0% | 23.0% | 15.3% | 33.3% | 23.9% | 33.7% |
| Sex abuse | | | 3.6% | | | | |
| Any child abuse | 15.5% | 5.6% | 24.6% | 17.9% | 34.4% | 25.4% | 34.8% |
| **Age 11-18** | | | | | | | |
| Child cruelty | 4.7% | 0.4% | 12.1% | 4.9% | 15.6% | 12.5% | 15.8% |
| Sex abuse | | | 5.4% | | | | |
| Any child abuse | 9.4% | 5.8% | 16.5% | 9.6% | 19.4% | 17.0% | 19.6% |

*Notes:* The table reports the prevalence of abuse in the estimation sample of 448 observations for Application 2. Letters in the column headers indicate the person who reported the measure of abuse: 'M' is for mothers, 'P' is for the mother's partner, and 'C' is for the child.

Table 5: Child abuse and age acceleration from the MEGA clock

| | SEM (1) | FA (2) | Wgt index (3) |
|---|---|---|---|
| Any child abuse (0-10) | 0.539** | 0.536** | 0.549** |
| | (0.251) | (0.255) | (0.252) |
| Any child abuse (11-18) | -0.007 | -0.134 | -0.143 |
| | (0.291) | (0.296) | (0.292) |
| Observations | 448 | 448 | 448 |
| Adjusted R-squared | . | 0.272 | 0.264 |

*Notes:* Standard errors in parentheses. 'SEM' stands for Structural Equation Modelling, 'FA' for Factor Analysis, and 'wgt index' for weighted index. All regressions control for mother's age at birth of the study child and binary indicators for mother's education, father's social class, and the child's gender, age, birth year, and birth order. $^{*}\ p < 0.1$, $^{**}\ p < 0.05$, $^{***}\ p < 0.01$.

Table 6: Child abuse and age acceleration from the MEGA clock: disaggregation

|  | SEM (1) | FA (2) | Wgt index (3) |
|---|---|---|---|
| Any child cruelty (0-10) | 0.546** | 0.543** | 0.559** |
|  | (0.254) | (0.259) | (0.255) |
| Any sex abuse (0-10) | 0.253 | 0.270 | 0.267 |
|  | (0.568) | (0.580) | (0.572) |
| Any child cruelty (11-18) | -0.054 | -0.228 | -0.214 |
|  | (0.314) | (0.319) | (0.315) |
| Any sex abuse (11-18) | 0.281 | 0.377 | 0.358 |
|  | (0.468) | (0.477) | (0.471) |
| Observations | 448 | 448 | 448 |
| Adjusted R-squared | . | 0.272 | 0.263 |

*Notes:* Standard errors in parentheses. 'SEM' stands for Structural Equation Modelling, 'FA' for Factor Analysis, and 'wgt index' for weighted index. All regressions control for mother's age at birth of the study child and binary indicators for mother's education, father's social class, and the child's gender, age, birth year, and birth order. $^{*}$ $p < 0.1$, $^{**}$ $p < 0.05$, $^{***}$ $p < 0.01$.

Table 7: The effect of delayed school entry on age acceleration at age 7

|  | (1) SEM | (2) FA | (3) Wgt Index |
|---|---|---|---|
| **Panel A: May - December** | | | |
| Treat | 0.317 | 0.542* | 0.579* |
|  | (0.235) | (0.321) | (0.313) |
| MoB | -0.142* | -0.192* | -0.198* |
|  | (0.074) | (0.104) | (0.101) |
| Treat * MoB | 0.056 | -0.017 | -0.002 |
|  | (0.117) | (0.142) | (0.139) |
| Observations | 597 | 597 | 597 |
| Adjusted R-squared |  | 0.116 | 0.130 |
| **Panel B: June - November** | | | |
| Treat | 0.328 | 0.440 | 0.448 |
|  | (0.252) | (0.398) | (0.392) |
| MoB | -0.091 | -0.105 | -0.107 |
|  | (0.109) | (0.167) | (0.164) |
| Treat * MoB | -0.134 | -0.174 | -0.131 |
|  | (0.163) | (0.226) | (0.223) |
| Observations | 446 | 446 | 446 |
| Adjusted R-squared |  | 0.083 | 0.094 |
| **Panel C: July - October** | | | |
| Treat | 0.891 | 1.158** | 1.114** |
|  | (0.587) | (0.541) | (0.531) |
| MoB | -0.457 | -0.647** | -0.625** |
|  | (0.365) | (0.310) | (0.304) |
| Treat * MoB | 0.054 | 0.376 | 0.448 |
|  | (0.341) | (0.444) | (0.436) |
| Observations | 307 | 307 | 307 |
| Adjusted R-squared |  | 0.069 | 0.076 |

*Notes:* Robust standard errors in parentheses. 'SEM' stands for Structural Equation Modelling, 'FA' for Factor Analysis, and 'wgt index' for weighted index. All regressions control for the child's age and dummies for the child's gender and birth year. * $p < 0.1$, ** $p < 0.05$, *** $p < 0.01$.

**Appendix A: Supplementary Figures and Tables**

Figure A1: Cross-rater, intertemporal correlations of child cruelty and sex abuse

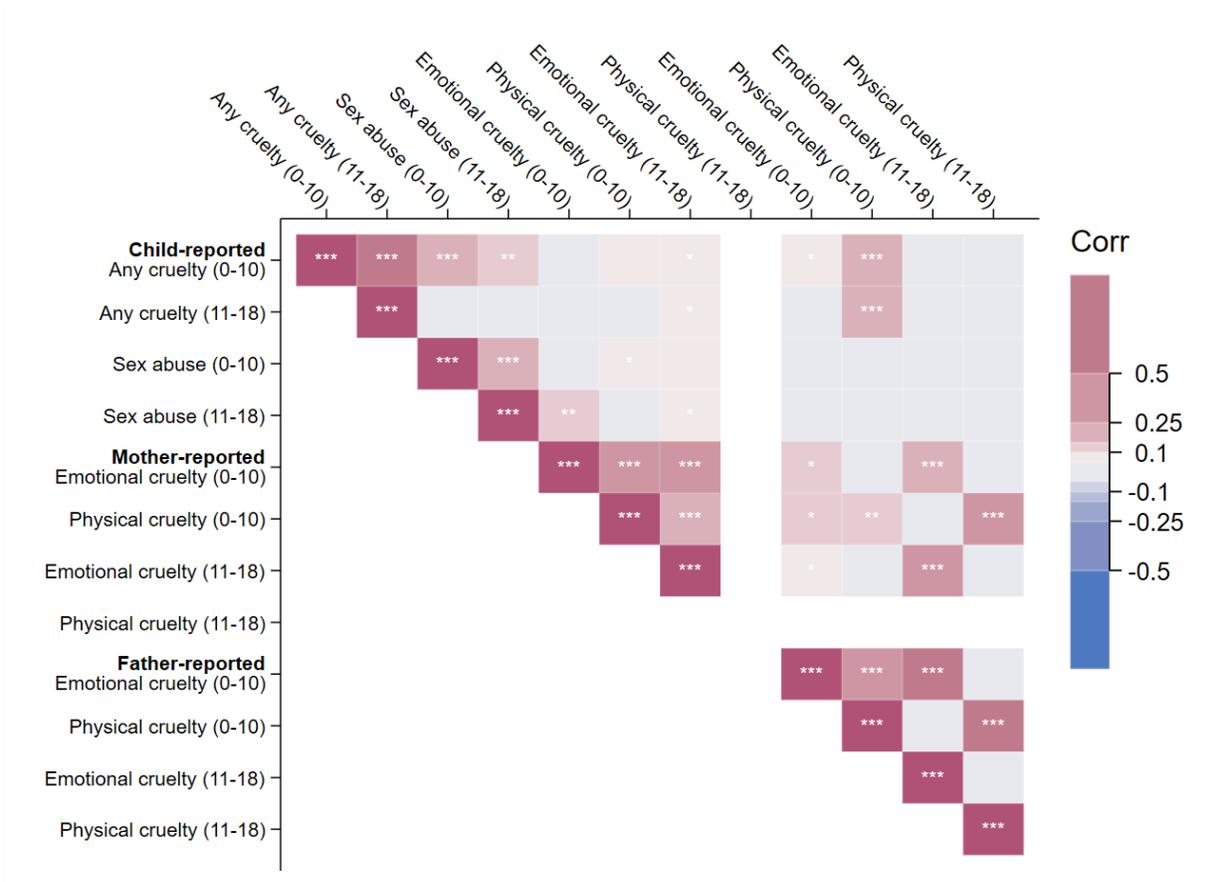

*Notes:* The graph replicates Figure 3 in the estimation sample of 448 participants from Application 2. The row and column corresponding to mother-reported physical cruelty between ages 11 and 18 are blank as there are no cases of mother-reported physical cruelty between child ages 11 and 18 in the estimation sample.

Figure A2: Chronological age and epigenetic age in four epigenetic clocks

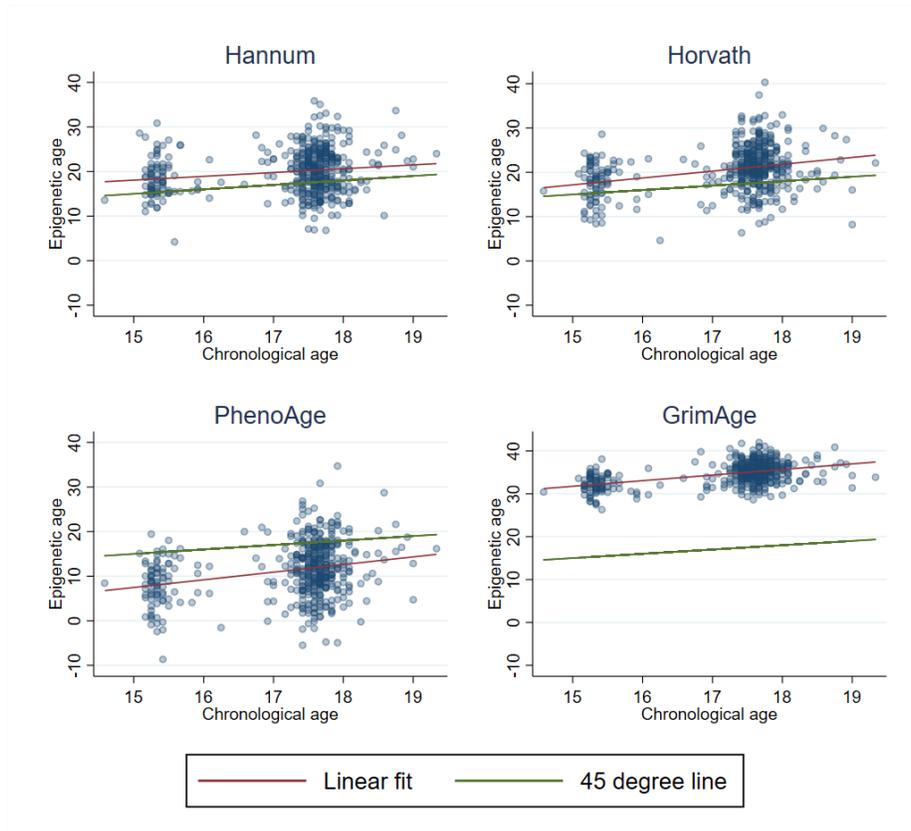

*Notes:* The graphs are scatterplots of epigenetic age against chronological age in the estimation sample of Application 2 (448 observations). Linear fits of the bivariate association are in red, while the 45-degree line is in green.

Figure A3: Chronological age and epigenetic age in the MEGA clocks

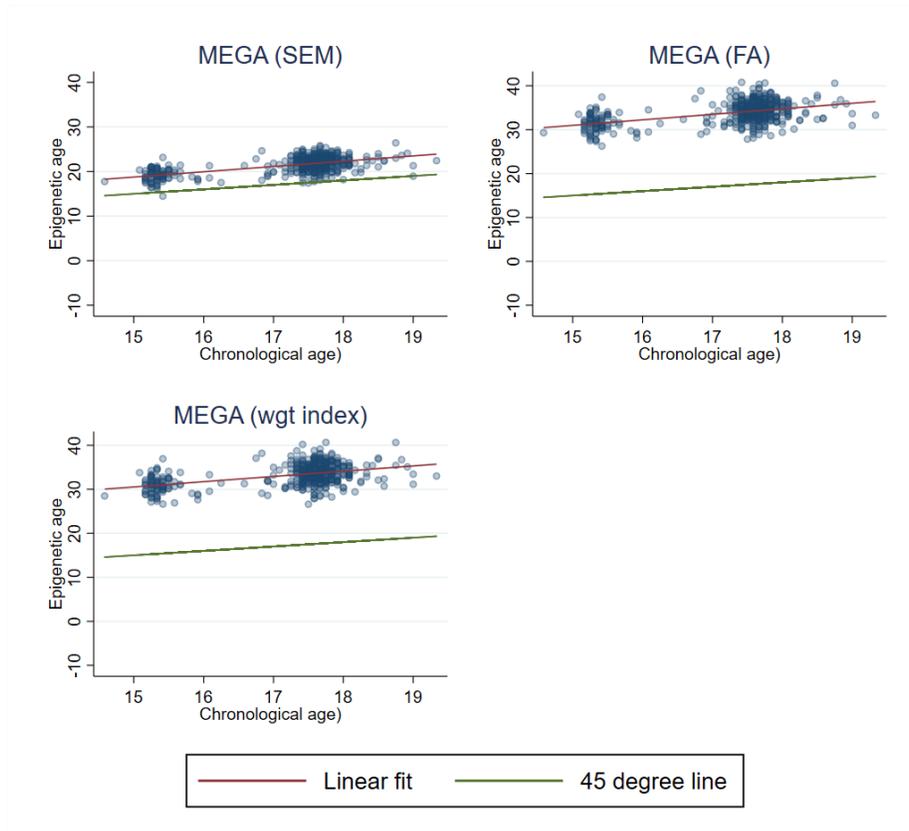

*Notes:* The graphs are scatterplots of epigenetic age against chronological age in the estimation sample of Application 2 (448 observations). 'SEM' stands for Structural Equation Modelling, 'FA' for Factor Analysis, and 'wgt index' for weighted index. Linear fits of the bivariate association are in red, while the 45-degree line is in green.

Figure A4: Correlation coefficients across clocks

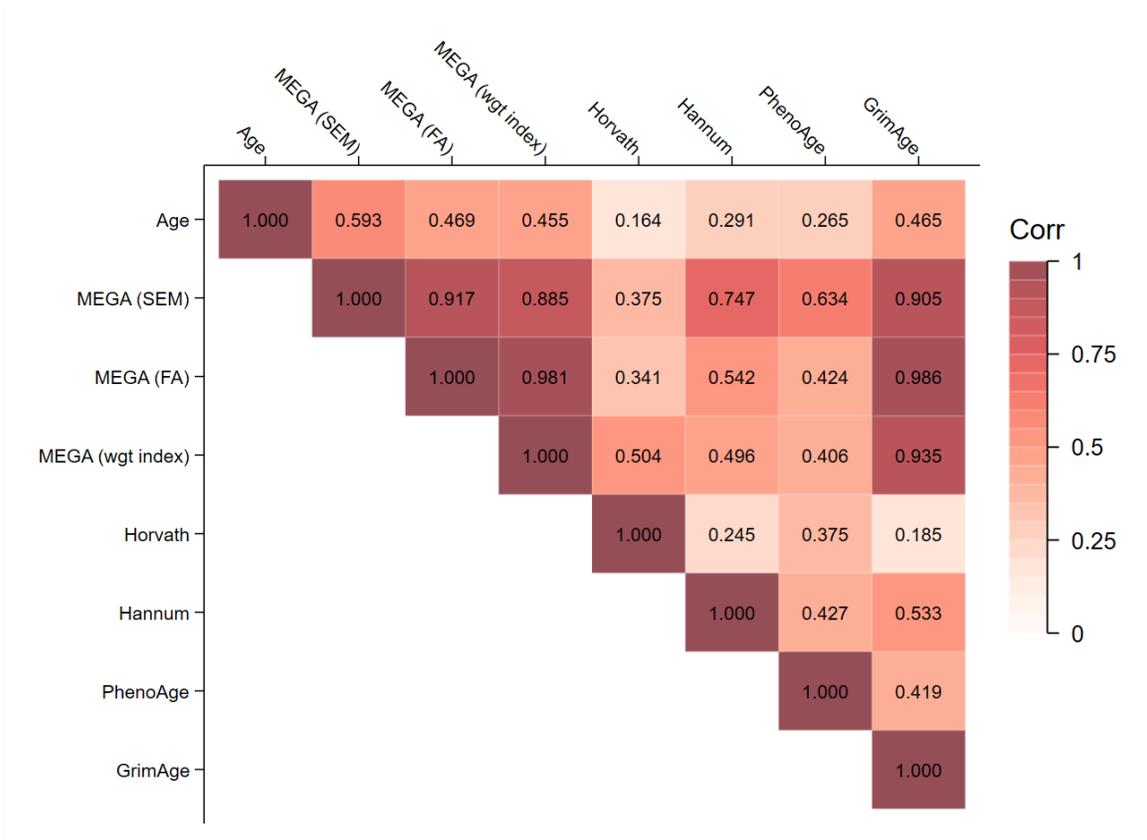

*Notes:* The correlation matrix refers to the estimation sample from Application 2 (448 observations). 'SEM' stands for Structural Equation Modelling, 'FA' for Factor Analysis, and 'wgt index' for weighted index.

Figure A5: Age acceleration, cognitive and socio-emotional skills, and early-adulthood outcomes

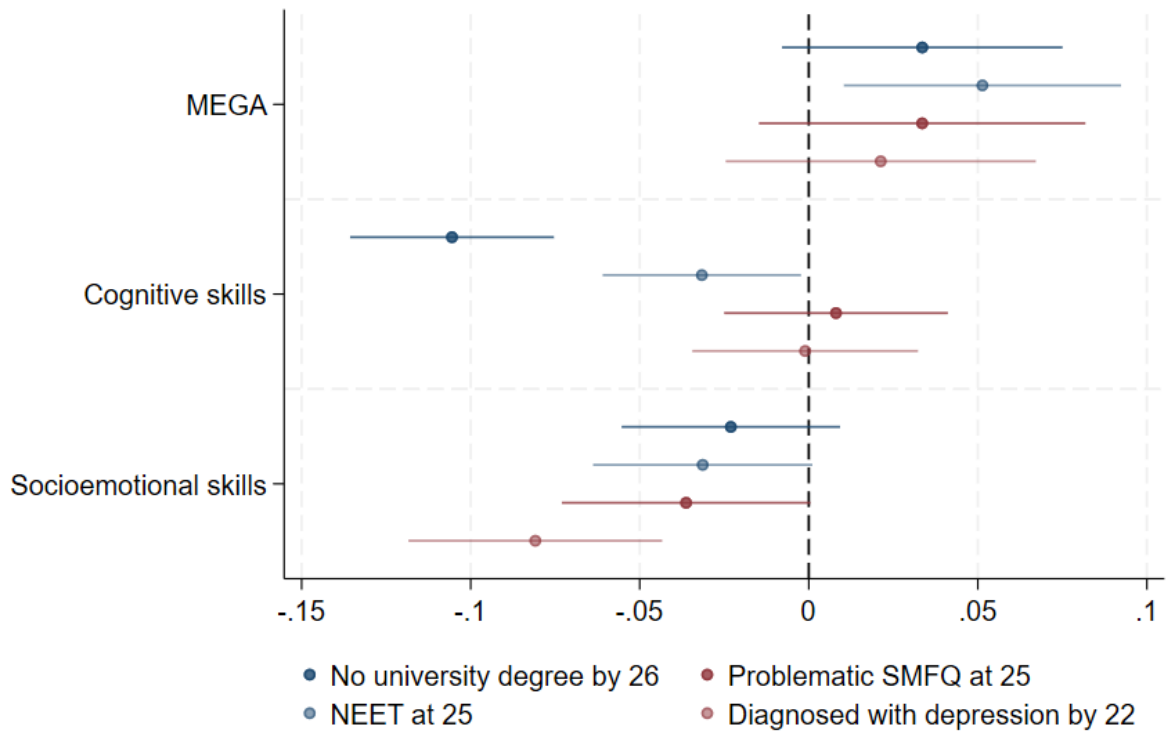

*Notes:* The figure replicates results in column (1) of Tables 2 and 3, additionally controlling for a cognitive and a socio-emotional skills factor. The MEGA clock and the cognitive and socio-emotional factors all come from a Structural Equation Model and are standardized in the figure, to enhance comparability. All regressions control for mother's age at birth of the study child and binary indicators for mother's education, father's social class, and the child's gender, birth year, and birth order. Horizontal spikes are for 90 percent confidence intervals.

Figure A6: Age acceleration by date of birth ($MEGA_{FA}$)

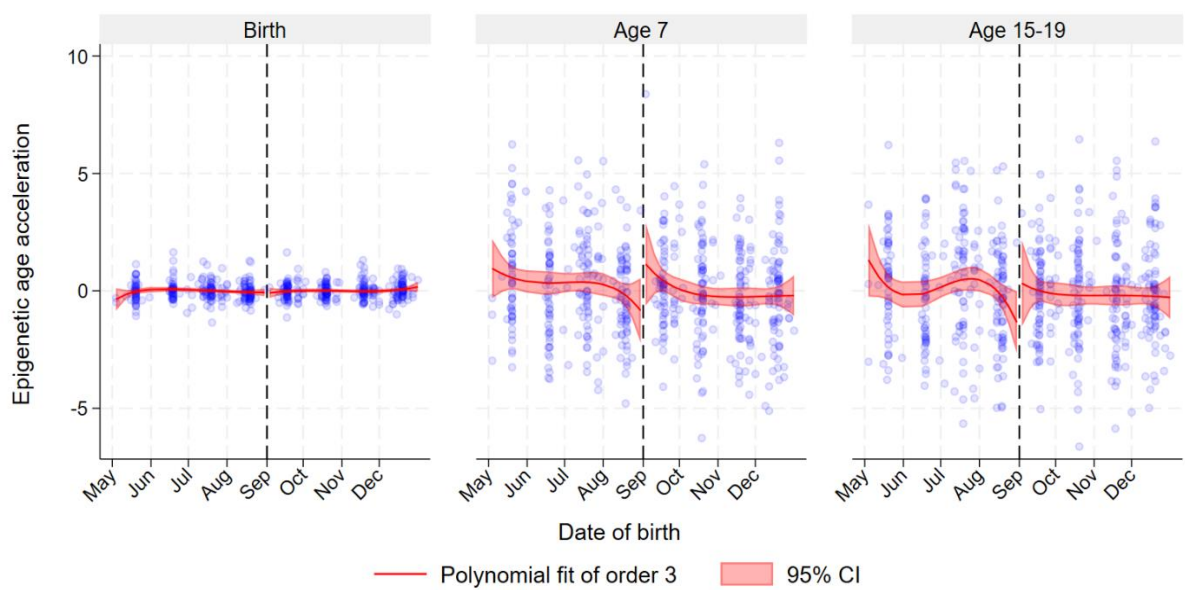

*Notes:* Scatter plot of epigenetic age acceleration, observed at the child's birth, age 7 and age 15-19, against date of birth. Epigenetic age acceleration is here computed as residuals from regressions of the MEGA (factor analysis) clock on chronological age, run for each of the three time points separately. To achieve greater granularity than we would by simply using month of birth, date of birth is here derived as the mid-point of a plausible date-of-birth interval, given by cross-referencing the child's birth month and birth year with the month and year of completion of the first child-based questionnaire, and the child's age in weeks at the time of completion. The vertical dashed line marks the 1ˢᵗ of September, the birthdate cut-off from which children's school entry is delayed by one academic year.

Figure A7: Age acceleration and early-adulthood cognitive outcomes: individual clocks

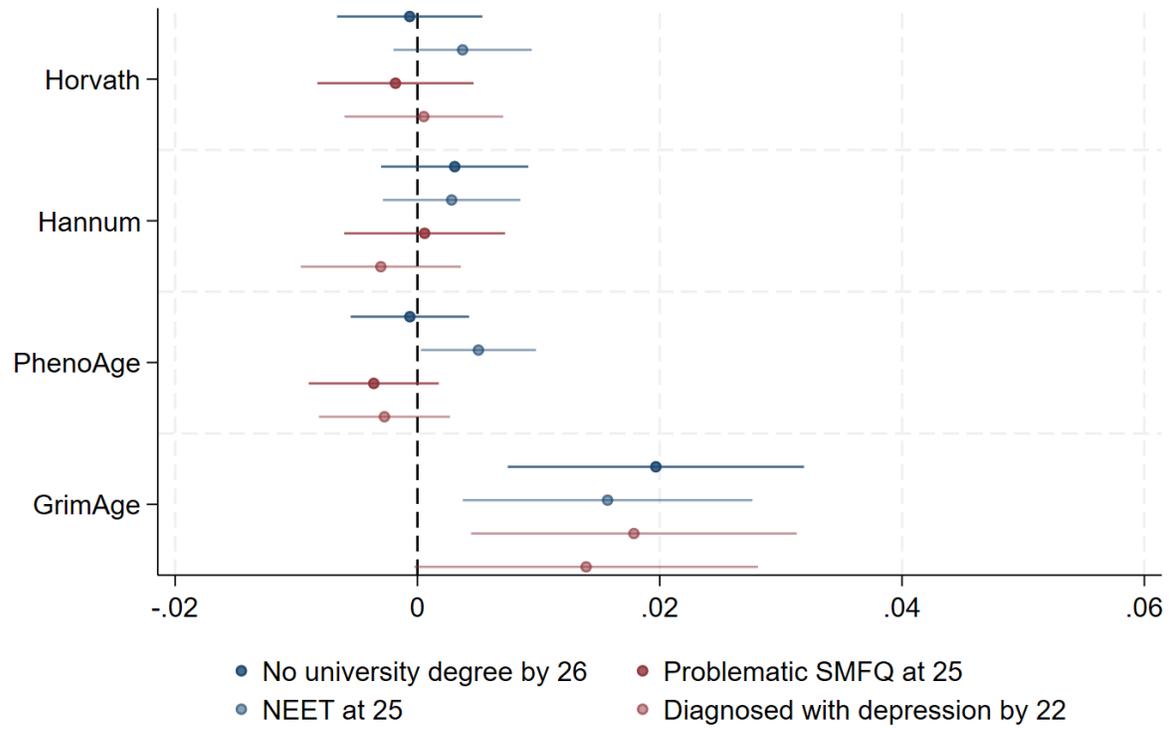

*Notes:* The figure replicates results from the even-numbered columns of Tables 2 and 3, using individual clocks as dependent variables instead of the MEGA clock. All regressions control for mother's age at birth of the study child and binary indicators for mother's education, father's social class, and the child's gender, birth year, birth order, and age 15 health outcomes (BMI, smoking, and drinking). Horizontal spikes are for 90 percent confidence intervals.

Figure A8: Child abuse (disaggregated) and age acceleration: individual clocks

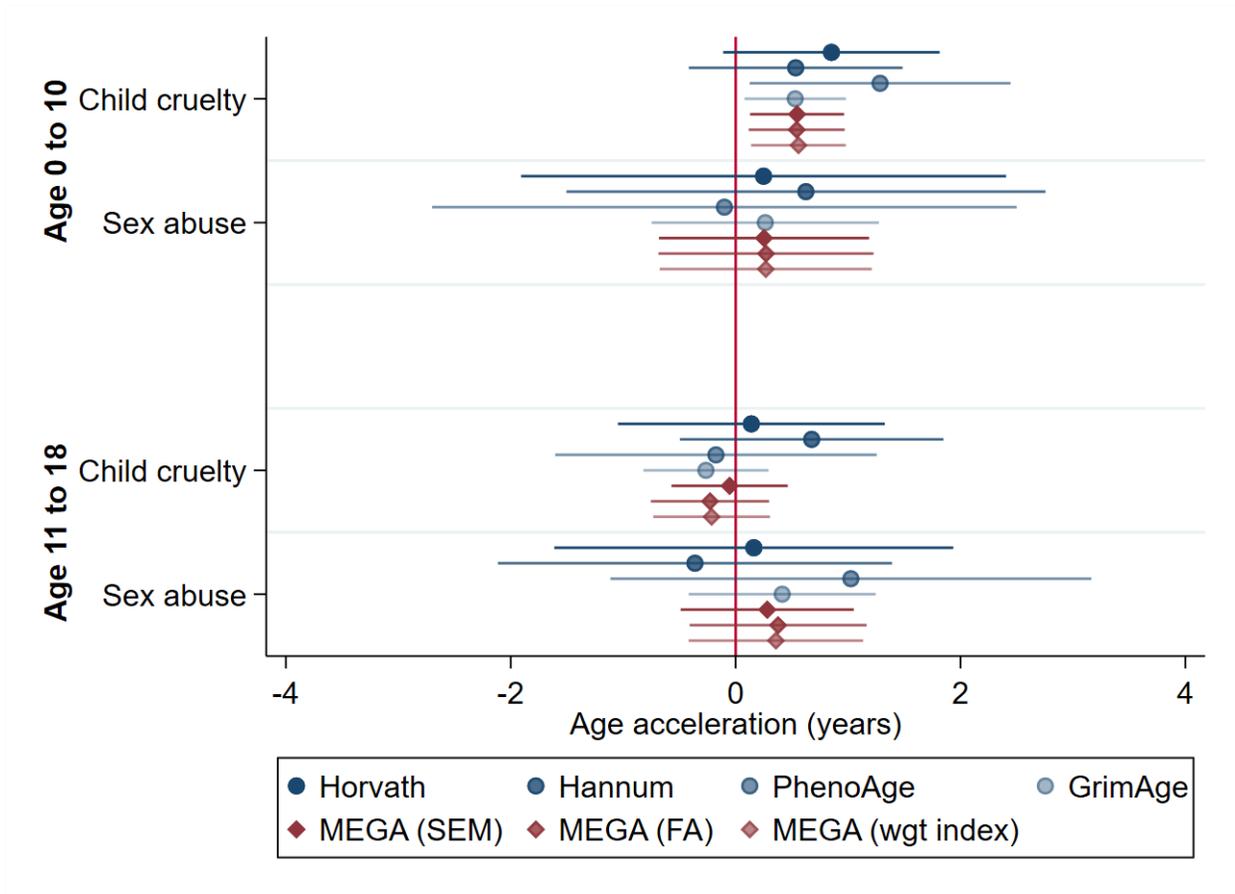

*Notes:* The blue dots in the figure replicate point estimates from Table 6 using the individual clocks as dependent variables, while the red dots report results for the MEGA clocks. 'SEM' stands for Structural Equation Modelling, 'FA' for Factor Analysis, and 'wgt index' for weighted index. All regressions control for mother's age at birth of the study child and binary indicators for mother's education, father's social class, and the child's gender, birth year, and birth order. Horizontal spikes are for 90 percent confidence intervals.

Figure A9: The effect of delayed school entry on age acceleration at age 7: individual clocks

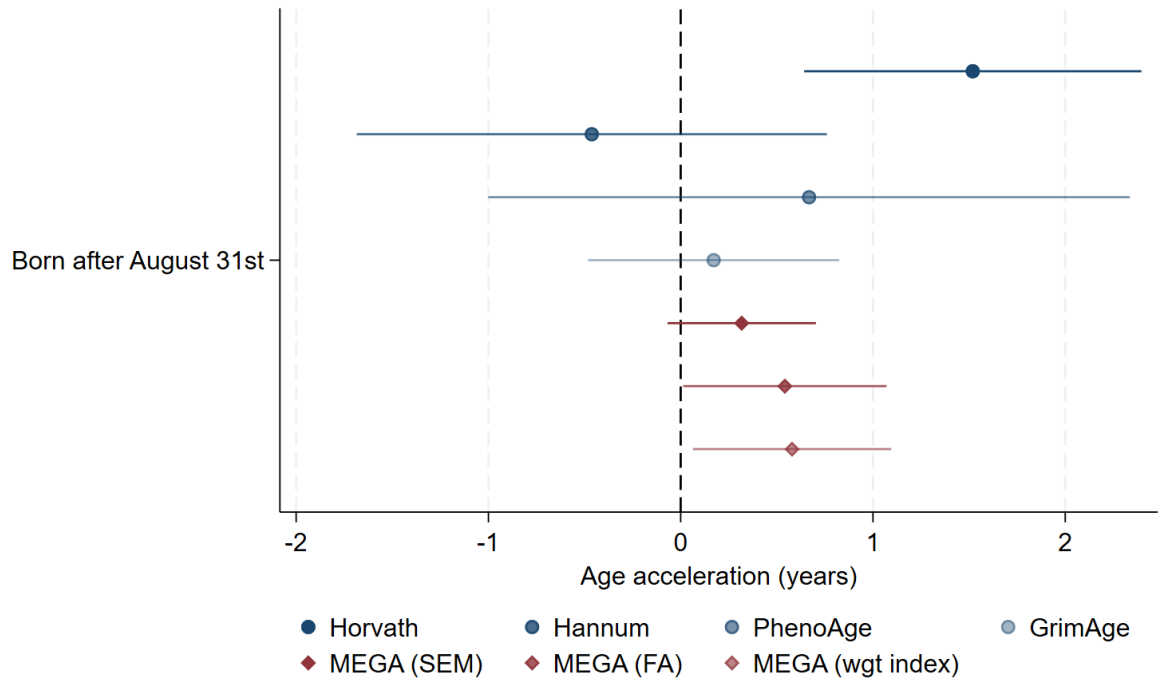

*Notes:* The blue dots in the figure replicate point estimates of the treatment dummy from Panel A of Table 7 using the individual clocks as dependent variables, while the red dots report results for the MEGA clocks. 'SEM' stands for Structural Equation Modelling, 'FA' for Factor Analysis, and 'wgt index' for weighted index. All regressions control for the child's gender, age and birth year. Horizontal spikes are for 90 percent confidence intervals.

Figure A10: Age acceleration and early-adulthood outcomes: leave-one-out MEGA

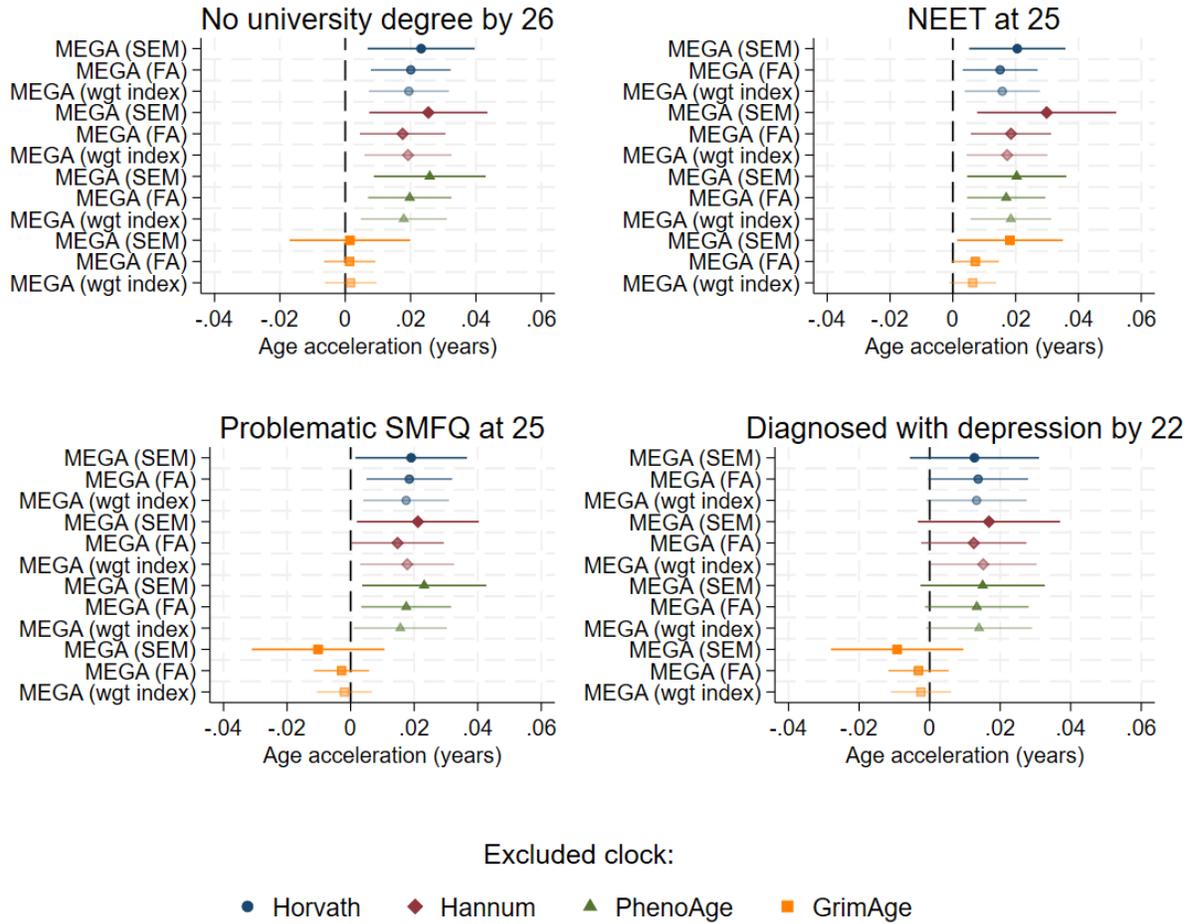

Figure A11: Child abuse and age acceleration: leave-one-out MEGA

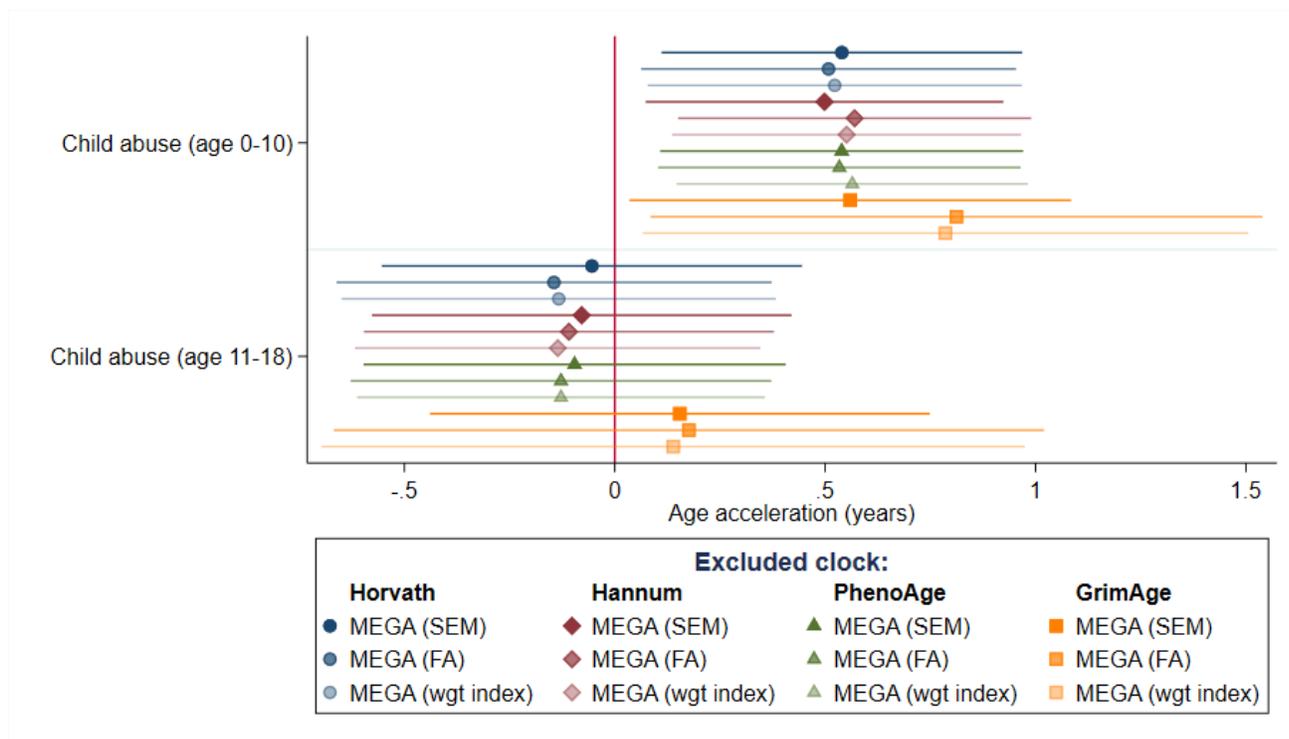

*Notes:* The figure replicates results from Table 5 using leave-one-out versions of the MEGA clock. 'SEM' stands for Structural Equation Modelling, 'FA' for Factor Analysis, and 'wgt index' for weighted index. All regressions control for mother's age at birth of the study child and binary indicators for mother's education, father's social class, and the child's gender and birth order. The child's birth year is also included as a control, except for the SEM model excluding the Hannum clock, where it is excluded in order to achieve model convergence. Horizontal spikes are for 90 percent confidence intervals.

Figure A12: Delayed school entry and age acceleration at age 7: leave-one-out MEGA

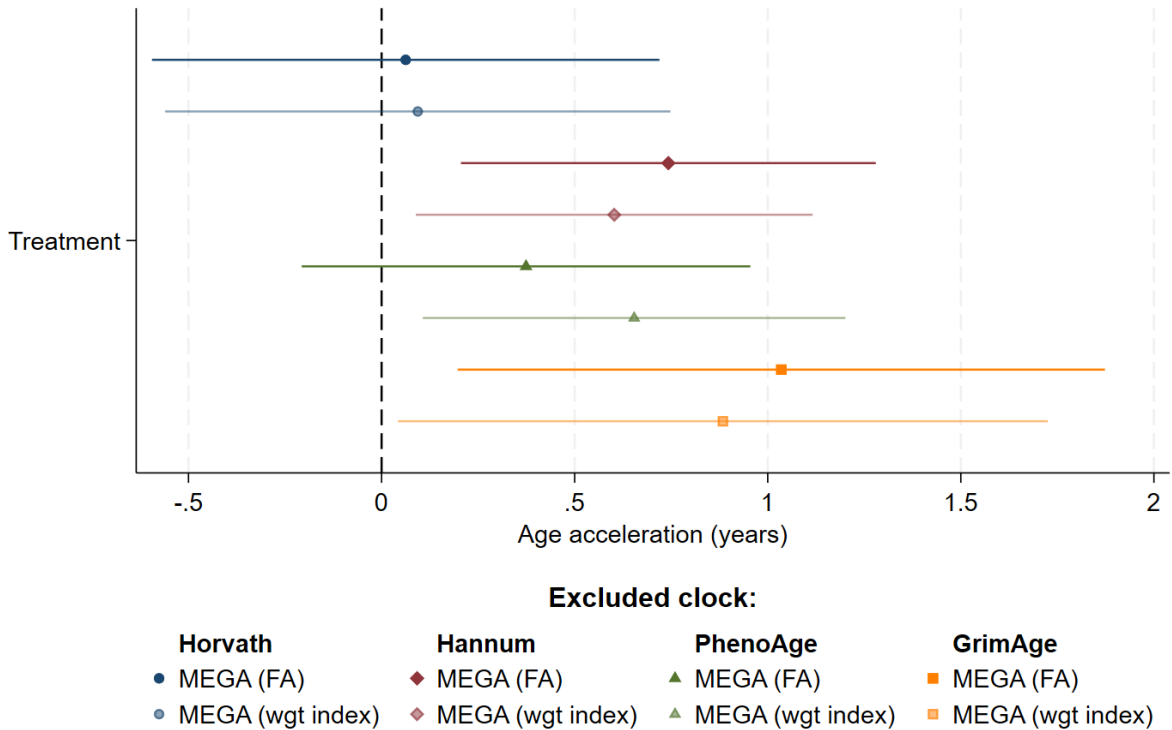

*Notes:* The figure replicates results from Panel A of Table 7 using leave-one-out versions of the MEGA clock. 'SEM' stands for Structural Equation Modelling, 'FA' for Factor Analysis, and 'wgt index' for weighted index. Dots are for the estimated coefficients of the treatment status variable. All regressions control for the child's age and binary indicators for the child's gender and birth year. Estimates relying on MEGA (SEM) are not illustrated here due to a lack of model convergence. Horizontal spikes are for 90 percent confidence intervals.

Figure A13: Age acceleration and blood cell counts by exposure

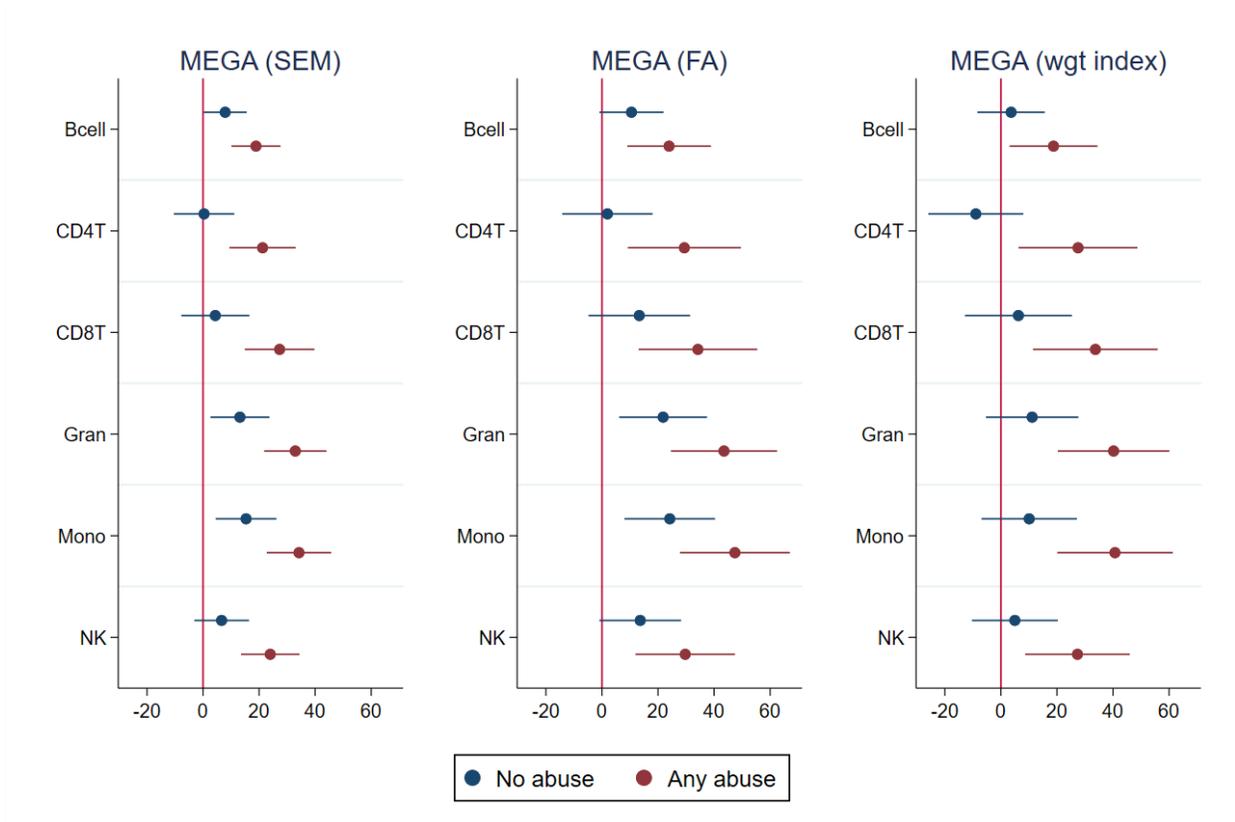

*Notes:* The figure plots associations between blood cell counts and the MEGA clocks in the estimation sample, from a linear regression model where the only control is age. 'SEM' stands for Structural Equation Modelling, 'FA' for Factor Analysis, and 'wgt index' for weighted index. Spikes are for 90 percent confidence intervals.

Table A1: Self-reported measures of child abuse in ALSPAC (age 22+)

| Variable label | Dichotomization |
|---|---|
| Frequency adult in family pushed, grabbed or shoved respondent | Happened at least 'sometimes' |
| Frequency adult in family smacked respondent for discipline | Happened at least 'sometimes' |
| Frequency adult in family punished respondent in a way that seemed cruel | Happened at least 'sometimes' |
| Frequency adult in family threatened to kick, punch, hit respondent with something that could hurt respondent or physically attack respondent in another way | Happened at least 'sometimes' |
| Frequency adult in family actually kicked, punched, hit respondent with something that could hurt respondent or physically attacked respondent in another way | Happened at least 'rarely' |
| Frequency adult in family hit respondent so hard it left bruises or marks | Happened at least 'rarely' |
| Respondent was touched in a sexual way by adult or older child, or was forced to touch adult or older child in a sexual way | Happened at least once |
| Adult or older child forced, or attempted to force, respondent into any sexual activity by threatening or holding respondent down or hurting respondent in some way | Happened at least once |

Table A2: Selection on observables of the estimation samples

| | Sample from Application: | | | | Differences | | |
|---|---|---|---|---|---|---|---|
| | No. 1 (1) | No. 2 (2) | No. 3 (3) | Full (4) | (4)-(1) | (4)-(2) | (4)-(3) |
| Female | 0.594 | 0.618 | 0.499 | 0.489 | -0.105*** | -0.129*** | -0.011 |
| | [0.492] | [0.486] | [0.500] | [0.500] | (0.021) | (0.024) | (0.021) |
| | *598* | *448* | *597* | *14997* | | | |
| Age | 17.177 | 17.228 | 7.457 | 17.123 | -0.054 | -0.105 | -0.003 |
| | [0.997] | [0.951] | [0.138] | [1.042] | (0.054) | (0.058) | (0.008) |
| | *598* | *448* | *597* | *925* | | | |
| Born in 1992 | 0.674 | 0.685 | 0.595 | 0.563 | -0.111*** | -0.122*** | -0.030 |
| | [0.469] | [0.465] | [0.491] | [0.496] | (0.021) | (0.024) | (0.021) |
| | *598* | *448* | *597* | *15468* | | | |
| First-born | 0.490 | 0.500 | 0.482 | 0.440 | -0.050* | -0.060* | -0.043* |
| | [0.500] | [0.501] | [0.500] | [0.496] | (0.021) | (0.024) | (0.021) |
| | *598* | *448* | *593* | *13320* | | | |
| Mother's age at birth | 29.843 | 29.946 | 29.668 | 27.989 | -1.854*** | -1.957*** | -1.677*** |
| | [4.319] | [4.307] | [4.462] | [4.969] | (0.206) | (0.238) | (0.207) |
| | *598* | *448* | *597* | *14023* | | | |
| Mother's education (ref: Lower-secondary) | | | | | | | |
| *Upper-secondary* | 0.296 | 0.292 | 0.295 | 0.182 | -0.114*** | -0.111*** | -0.113*** |
| | [0.457] | [0.455] | [0.456] | [0.386] | (0.016) | (0.019) | (0.016) |
| | *598* | *448* | *597* | *15612* | | | |
| *Post-secondary* | 0.258 | 0.277 | 0.238 | 0.104 | -0.154*** | -0.173*** | -0.134*** |
| | [0.438] | [0.448] | [0.426] | [0.305] | (0.013) | (0.015) | (0.013) |
| | *598* | *448* | *597* | *15612* | | | |
| Father's social class (ref.: Professionals) | | | | | | | |
| *Non-manual* | 0.467 | 0.482 | 0.447 | 0.316 | -0.151*** | -0.166*** | -0.131*** |
| | [0.499] | [0.500] | [0.498] | [0.465] | (0.019) | (0.022) | (0.019) |
| | *598* | *448* | *597* | *15584* | | | |
| *Manual* | 0.289 | 0.259 | 0.317 | 0.311 | 0.022 | 0.052* | -0.006 |
| | [0.454] | [0.439] | [0.466] | [0.463] | (0.019) | (0.022) | (0.019) |
| | *598* | *448* | *597* | *15584* | | | |

*Notes:* The table plots means of covariates and their differences across the estimation samples of the three empirical applications and the largest ALSPAC sample in which each covariate is available. Standard deviations in brackets and standard errors in parentheses. Sample sizes are indicated in italics below standard deviations. $^* p < 0.1$, $^{**} p < 0.05$, $^{***} p < 0.01$.

Table A3: Factor analysis results for $MEGA_{FA}$

| | Application 1 (age 15-19, N=598) | | Application 2 (age 15-19, N=448) | | Application 3 (age 7, N=597) | |
|---|---|---|---|---|---|---|
| | Factor loadings | Uniqueness | Factor loadings | Uniqueness | Factor loadings | Uniqueness |
| Horvath | 0.422 | 0.822 | 0.410 | 0.832 | 0.500 | 0.750 |
| Hannum | 0.659 | 0.565 | 0.662 | 0.562 | 0.603 | 0.636 |
| PhenoAge | 0.652 | 0.575 | 0.633 | 0.599 | 0.701 | 0.509 |
| GrimAge | 0.623 | 0.612 | 0.639 | 0.592 | 0.527 | 0.722 |

*Notes:* Values are loadings obtained from factor analysis of the four epigenetic clocks, run on the largest estimation sample from each application. All applications support a unifactoral model, with only one factor displaying eigenvalue greater than one. For Application 1, this is the sample of 598 observations for which all controls, all clocks and at least one of the four adult outcomes are available. For Application 2, this is the estimation sample of 448 observations. Last, for Application 3, we rely on the May-December estimation sample of 597 observations at child age 7.

Table A4: Child abuse and age acceleration from the MEGA clock: sensitivity to the rater of abuse

| | M (1) | P (2) | C (3) | MP (4) | CM (5) | CP (6) | CMP (7) |
|---|---|---|---|---|---|---|---|
| **A. SEM** | | | | | | | |
| Any abuse (0-10) | . | 0.528 | 0.488$^*$ | 0.294 | 0.489$^*$ | 0.558$^{**}$ | 0.539$^{**}$ |
| | . | (0.458) | (0.277) | (0.283) | (0.251) | (0.273) | (0.251) |
| Any abuse (11-18) | . | 0.447 | -0.010 | 0.321 | 0.024 | -0.007 | -0.007 |
| | . | (0.450) | (0.315) | (0.362) | (0.292) | (0.311) | (0.291) |
| Observations | 448 | 448 | 448 | 448 | 448 | 448 | 448 |
| **B. FA** | | | | | | | |
| Any abuse (0-10) | 0.408 | 0.400 | 0.308 | 0.427 | 0.470$^*$ | 0.382 | 0.536$^{**}$ |
| | (0.295) | (0.466) | (0.284) | (0.286) | (0.256) | (0.280) | (0.255) |
| Any abuse (11-18) | 0.367 | 0.482 | -0.035 | 0.327 | -0.084 | -0.058 | -0.134 |
| | (0.372) | (0.457) | (0.324) | (0.367) | (0.297) | (0.320) | (0.296) |
| Observations | 448 | 448 | 448 | 448 | 448 | 448 | 448 |
| Adjusted R-squared | 0.174 | 0.177 | 0.181 | 0.176 | 0.181 | 0.186 | 0.184 |
| **C. Wgt index** | | | | | | | |
| Any abuse (0-10) | 0.468 | 0.392 | 0.315 | 0.489$^*$ | 0.480$^*$ | 0.386 | 0.549$^{**}$ |
| | (0.290) | (0.460) | (0.280) | (0.282) | (0.252) | (0.277) | (0.252) |
| Any abuse (11-18) | 0.348 | 0.443 | -0.033 | 0.299 | -0.086 | -0.065 | -0.143 |
| | (0.367) | (0.451) | (0.319) | (0.362) | (0.293) | (0.316) | (0.292) |
| Observations | 448 | 448 | 448 | 448 | 448 | 448 | 448 |
| Adjusted R-squared | 0.263 | 0.259 | 0.258 | 0.264 | 0.262 | 0.259 | 0.264 |

*Notes:* Standard errors in parentheses. The dependent variable is the MEGA clock age acceleration, computed with Structural Equation Modelling in panel A, with factor analysis in panel B and with the weighted index in panel C. Letters in the column headers indicate the person who reported the measure of child cruelty used in the definition of 'Any abuse': 'M' is for mothers, 'P' is for the mother's partner, and 'C' is for the child. All regressions control for mother's age at birth of the study child and binary indicators for mother's education, father's social class, and the child's gender, birth year, and birth order. $^*$ $p < 0.1$, $^{**}$ $p < 0.05$, $^{***}$ $p < 0.01$.

Table A5: The effect of delayed school entry on age acceleration at age 15-19

| | (1) SEM | (2) FA | (3) Wgt Index |
|---|---|---|---|
| **Panel A: May - December** | | | |
| Treat | 0.132 | 0.108 | 0.174 |
| | (0.343) | (0.356) | (0.353) |
| | | | |
| MoB | 0.009 | -0.016 | -0.047 |
| | (0.114) | (0.118) | (0.117) |
| | | | |
| Treat * MoB | -0.088 | -0.087 | -0.044 |
| | (0.153) | (0.159) | (0.157) |
| Observations | 600 | 600 | 600 |
| Adjusted R-squared | | 0.291 | 0.280 |
| **Panel B: June - November** | | | |
| Treat | . | 0.146 | 0.268 |
| | | (0.431) | (0.427) |
| | | | |
| MoB | . | -0.003 | -0.039 |
| | | (0.188) | (0.187) |
| | | | |
| Treat * MoB | . | -0.124 | -0.108 |
| | | (0.254) | (0.252) |
| Observations | | 454 | 454 |
| Adjusted R-squared | | 0.302 | 0.286 |
| **Panel C: July - October** | | | |
| Treat | . | 0.234 | 0.356 |
| | | (0.597) | (0.592) |
| | | | |
| MoB | . | -0.745** | -0.820** |
| | | (0.344) | (0.341) |
| | | | |
| Treat * MoB | . | 0.739 | 0.803* |
| | | (0.490) | (0.486) |
| Observations | | 317 | 317 |
| Adjusted R-squared | | 0.314 | 0.299 |

*Notes:* Standard errors in parentheses. 'SEM' stands for Structural Equation Modelling, 'FA' for Factor Analysis, and 'wgt index' for weighted index. SEM estimates are not provided in Panel B and C since the estimating algorithm did not autonomously converge, likely due to the small sample size. Omitting some controls or changing the starting point of the algorithm would allow for convergence but at the expense of clarity and comparability. All regressions control for the child's age and dummies for the child's gender and birth year. $^*$ $p < 0.1$, $^{**}$ $p < 0.05$, $^{***}$ $p < 0.01$.

Table A6: Child abuse and age acceleration from the MEGA clock (controlling for cell type counts)

| | SEM (1) | FA (2) | Wgt index (3) |
|---|---|---|---|
| Any child abuse (0-10) | . | 0.305 (0.192) | 0.340[*] (0.203) |
| Any child abuse (11-18) | . | -0.360 (0.222) | -0.355 (0.235) |
| Observations | | 448 | 448 |
| Adjusted R-squared | | 0.595 | 0.525 |

*Notes:* Standard errors in parentheses. 'SEM' stands for Structural Equation Modelling, 'FA' for Factor Analysis, and 'wgt index' for weighted index. SEM estimates are not provided since the estimating algorithm did not reliably converge, likely due to the small sample size. All regressions control for mother's age at birth of the study child and binary indicators for mother's education, father's social class, and the child's gender, birth year, and birth order.[*] $p < 0.1$, [**] $p < 0.05$, [***] $p < 0.01$.